\documentclass[aps,prd,10pt,superscriptaddress,amsfonts,amssymb,amsmath,showpacs]{revtex4-2}
\usepackage{bm}
\usepackage{latexsym}
\usepackage[utf8]{inputenc}
\usepackage{graphicx}
\usepackage{palatino}
\usepackage{mathpazo}
\usepackage{textcomp}
\usepackage{float}
\usepackage{booktabs}
\usepackage{dcolumn}
\usepackage{ragged2e}
\usepackage{xcolor}
\usepackage{hyperref}
\hypersetup{colorlinks,citecolor=blue,urlcolor=blue}
\usepackage{amsmath}
\usepackage{orcidlink}
\usepackage{epsfig}
\usepackage{cleveref}
\usepackage{caption}
\usepackage{subcaption}
\usepackage{commath}

\allowdisplaybreaks[1]

\begin{document}

\color{black}    

\title{Constraints on Cosmological Parameters to Ease $H_0$ Tension in $f(R, \mathcal{L}_m)$ Gravity} 
\title{Easing $H_0$ tension with Constrained Cosmological Parameters in $f(R, \mathcal{L}_m)$ Gravity} 
\author{Y. Kalpana Devi\orcidlink{0009-0001-2686-7281}}
\email{kalpanayengkhom123@gmail.com}
\affiliation{Department of Mathematics,
Birla Institute of Technology and Science-Pilani, Hyderabad Campus, Jawahar Nagar, Kapra Mandal, Medchal District, Telangana 500078, India.}

\author{B. Mishra\orcidlink{0000-0001-5527-3565}}
\email{bivu@hyderabad.bits-pilani.ac.in}
\affiliation{Department of Mathematics,
Birla Institute of Technology and Science-Pilani, Hyderabad Campus, Jawahar Nagar, Kapra Mandal, Medchal District, Telangana 500078, India.}

\author{Francisco Tello-Ortiz \orcidlink{0000-0002-7104-5746}}
\email{francisco.tello@ufrontera.cl}
\affiliation{Departamento de Ciencias F\'isicas, Universidad de La Frontera, Casilla 54-D, 4811186 Temuco, Chile.}

\date{\today}
\begin{abstract}

\textbf{Abstract}: 
We constrain two $f(R,\mathcal{L}_{m})$ gravity models formulated through power-law and exponential curvature--matter couplings to evaluate their late-time dynamics and capacity to alleviate the $H_{0}$ tension. Using Markov Chain Monte Carlo sampling within the Cobaya framework, we confront both models against Cosmic Chronometers (CC), Pantheon+ Type Ia Supernovae ($\text{SN}^{+}$), and DESI-DR2 Baryon Acoustic Oscillation (BAO) data under uniform and local SH0ES(2022) Gaussian $H_{0}$ priors, benchmarking performance against flat $\Lambda\text{CDM}$ via $\Delta\text{AIC}$, $\Delta\text{BIC}$, and $\Delta\text{DIC}$. In the unanchored joint analysis ($\text{CC}+\text{SN}^{+}+\text{BAO}$), both extensions are statistically favored over $\Lambda\text{CDM}$ ($\Delta\text{BIC} = -6.60$ and $-2.91$, respectively) and drive a smooth transition to late-time acceleration at $z_{t} \approx 0.62\text{--}0.67$ for the power-law model and $z_{t} \approx 0.56\text{--}0.60$ for the exponential model. Relative to the unanchored $\Lambda\text{CDM}$ constraint of $H_{0} = 68.24 \pm 0.39\,\mathrm{km\,s^{-1}\,Mpc^{-1}}$ which is around $4.3\sigma$ tension with SH0ES, the power-law model yields $H_{0} = 67.00 \pm 1.04\,\mathrm{km\,s^{-1}\,Mpc^{-1}}$ ($4.1\sigma$), whereas the exponential model modestly elevates $H_{0}$ to $69.46 \pm 0.70\,\mathrm{km\,s^{-1}\,Mpc^{-1}}$, easing the tension to $2.9\sigma$ without local calibration. Under the SH0ES prior, the power-law model exhibits dataset-dependent phantom excursions ($q \to -1.21$, $\omega_{\mathrm{eff}} \to -1.14$ in $\text{CC}+H_{0}$) and incurs a strong Bayesian penalty in the full joint fit ($\Delta\text{BIC} = +7.42$). Conversely, the exponential model maintains strict quintessence-like stability, converging smoothly to the de Sitter attractor ($q \to -1$, $\omega_{\mathrm{eff}} \to -1$) while reducing the SH0ES discrepancy to $1.8\sigma$ ($H_{0} = 70.60 \pm 0.86\,\mathrm{km\,s^{-1}\,Mpc^{-1}}$) with positive Akaike support ($\Delta\text{AIC} = -4.29$) and only a mild Bayesian complexity penalty ($\Delta\text{BIC} = +1.17$).
\\

\end{abstract}

\maketitle
\textbf{Keywords}: Curvature-Matter coupled gravity, Hubble tension, Late-time behavior, Deceleration parameter.

\section{Introduction}\label{I}

\noindent The $\Lambda$-Cold Dark Matter ($\Lambda$CDM) cosmological model, grounded in General Relativity (GR) with a cosmological constant, provides a remarkably accurate framework for describing the large-scale structure and evolution of the Universe \cite{einstein1920relativity, lorentz2026principle, Carroll_2001, Weinberg:2008}. Despite its empirical success, the model faces severe theoretical challenges, including the cosmological constant problem and the coincidence problem \cite{Weinberg:1988cp, Peebles_2003}. Modern cosmology is further confronted by persistent observational tensions, most prominently the $4\sigma$--$6\sigma$ discrepancy in the Hubble constant ($H_0$) inferred from early-universe Cosmic Microwave Background (CMB) measurements\cite{Bennett_2003, Spergel_2003, Riess_2019, Verde_2019} versus local distance ladder observations \cite{Perlmutter_1997, Perlmutter_1998, Perlmutter_1999, Riess_2004, Riess_2007, efstathiou2024challengeslambdacdmcosmology}. Adding to this, $S_8$ tension in large-scale structure clustering growth \cite{2017ARA&A..55..343B, 2017ARA&A..55..343B, DiValentino_2021, YADAV2026100453, 10.1007/978-3-031-85527-6_13} poses another challenge. These empirical crises strongly indicate new physics beyond the Standard Model and led to the investigation of modified gravity frameworks as potential alternatives to dark energy \cite{Clifton_2012, Nojiri_2017, Koyama_2016}. Beyond the $\Lambda$CDM model, higher-order modified theories of gravity such as $f(R)$ theory \cite{Carroll_2004, De_Felice_2010, Sotiriou_2010}, scalar–tensor theory \cite{Naruko_2016, Riess_1998,  faraoni2004cosmology}, and teleparallel extensions with torsional formulations of $f(T)$ gravity \cite{PhysRevD.92.104042, Cai:2015emx} seek to explain late-time cosmic acceleration without invoking an ad hoc dark energy fluid by changing the gravitational sector. Their relevance is strengthened by the fact that cosmological tensions may leave characteristic signatures in both the background expansion and the growth of the structure. However, resolving the discrepancies of both $H_0$ and $S_8$ remain a stringent challenge with simple extensions \cite{Abdalla_2022, boiza2025addressingh0s8tensions}.\\

\noindent The $f(R,\mathcal{L}_m)$ gravity \cite{Harko_2010b}, in which the gravitational Lagrangian is an arbitrary function of both the Ricci scalar $R$ and the matter Lagrangian density $\mathcal{L}_m$ is another geometry-matter coupling gravity. This explicit non-minimal curvature-matter coupling induces an extra force acting on massive particles, leading to non-geodesic motion and violating the equivalence principle \cite{Harko_2010, Harko_2012, Harko_2014, Wang_2012}. Theories of $f(R,\mathcal{L}_m)$ gravity offer rich cosmological dynamics, capable of reproducing late-time acceleration, mimicking quintessence-like behavior, and modifying energy conditions without demanding exotic matter components \cite{Avelino:2018rsb, gonçalves2024causalityviolationfrlmtgravity}.
Within this broader class, Gaussian-process reconstruction offers a non-parametric means of inferring the expansion history and its derivatives from observations without imposing a predetermined functional form, thereby reducing dependence on a specific cosmological parameterization \cite{Seikel_2012}. This approach is particularly suitable for combining cosmic-chronometer (CC) measurements with Type Ia supernovae and baryon acoustic oscillations (BAO) data, as it allows data-driven reconstruction of $H(z)$ and related cosmological quantities \cite{Seikel2013, Moresco_2016, Zhang_2016}. The cosmological viability of Starobinsky-type and power-law forms of $f(R,\mathcal{L}_m)$ gravity was studied by constraining their model and cosmological parameters using cosmological datasets \cite{DEVI2024101640}. The Gaussian-process reconstruction has been employed to obtain the late-time expansion history in a model-independent manner for power-law and exponential $f(R,\mathcal{L}_m)$ models through a dynamical-system analysis \cite{devi2025late}. The scalar field incorporated exponential model employing central manifold theory shows the stability of critical points \cite{DEVI2026170597}.\\

\noindent Here, we are motivated to investigate two functional forms of $f(R,\mathcal{L}_m)$ gravitational models using CC, SN$^+$ SNe Ia, and recent DESI-DR2 BAO measurements.  We perform this comparative analysis to determine whether the effect of non-minimal coupling significantly relax existing observational tensions by altering the background expansion history or not. In particular, we examine how imposing a uniform  $H_0$ prior, relative to its omission, shifts the posterior constraints, uncertainties, and relative consistency of the $f(R,\mathcal{L}_m)$ models, when applied to the baseline Pantheon and CC samples and their combination with DESI-DR2 BAO data. The analysis utilizes the model-independent Gaussian-process (GP) reconstruction of the cosmic expansion history obtained with GaPP (Gaussian Processes in Python)\cite{Seikel_2012, Moresco_2016, Riess_2018, Zhang_2016,  devi2025late} and then analyzes the reconstructed expansion history within the specific $f(R,\mathcal{L}_m)$ forms. More specifically, we assess their parameter constraints and cosmological viability. We evaluated whether these specific gravitational modifications retain sufficient degrees of freedom to capture the underlying expansion data while simultaneously addressing the reported tension in the local Hubble rate \cite{Wolf_2025, Di_Valentino_2026}. This comparative analysis probes the relative impact of local and high-redshift observations on the allowed model parameter space and tests whether the geometric matter coupling can alleviate existing cosmological tensions \cite{Kumar_Aluri_2023, BERNARDO2023101213}. In particular, we examine how the inclusion of a local $H_0$ prior modifies the posterior distributions and whether it introduces significant shifts in the inferred cosmological parameters. We further examine the parameter shifts induced by the DESI DR2 BAO measurements to test the robustness of the power and exponential-law $f(R,\mathcal{L}_m)$ models against recent high-precision distance--redshift constraints and their implications for departures from $\Lambda$CDM \cite{tr6y-kpc6}.\\

\noindent We organize the paper as follows: Section~\ref{I}  provides an introduction that focuses on the observational tensions confronting $\Lambda$CDM and the motivation for $f(R,\mathcal{L}_m)$ gravity. In Section~\ref{II}, we discuss the theoretical foundation of $f(R,\mathcal{L}_m)$ gravity and the two forms of $f(R,\mathcal{L}_m)$ such as the power and exponential-law  formulations. In addition, the corresponding field equations and background expressions for $H(z)$, $q(z)$, and $w_{\mathrm{eff}}(z)$ are derived for both models. In Section~\ref{III}, we describe the observational datasets—comprising CC, Pantheon$^{+}$ SNe Ia, and DESI DR2 BAO—alongside the $MCMC$ setup and model-selection metrics ($\chi^2$, AIC, BIC, DIC). In Section~\ref{IV}, we report the posterior parameter constraints, contour plots, and statistical comparison metrics ($\Delta\mathrm{AIC}$, $\Delta\mathrm{BIC}$, $\Delta\mathrm{DIC}$) relative to $\Lambda$CDM for both prior choices. In Section~\ref{V}, we analyze the evolutionary trajectories of $q(z)$ and $w_{\mathrm{eff}}(z)$ to evaluate cosmic acceleration and dark-energy dynamics. Finally, in Section~\ref{VI}, we synthesize our primary findings, while the Appendix provides supplementary baseline $\Lambda$CDM posterior plots and model-selection tables.

\section{The Theoretical Framework}\label{II}
The action of $f(R,\mathcal{L}_m)$ gravity \cite{Harko_2010b} is defined as,
\begin{equation}\label{action}
S=\int f(R,\mathcal{L}_m) \sqrt{-g} d^4x .
\end{equation}
We take $8\pi G=1$. The function $f(R,\mathcal{L}_m)$ is a function of Ricci scalar $R$ and matter Lagrangian $\mathcal{L}_m$ the determinant of the metric tensor $g_{\mu\nu}$ becomes $\sqrt{-g}$.  
Now, one can obtain the field equations of $f(R,\mathcal{L}_m)$ gravity by varying the action principle Eq. \eqref{action} with respect to the metric tensor, 

\begin{equation}\label{Eq:field1}
R_{\mu\nu}f_R+(g_{\mu\nu}\Box-\nabla_\mu\nabla_\nu )f_R-\frac{f}{2}g_{\mu\nu}
    =\frac{1}{2}f_{\mathcal{L}_m}(T_{\mu\nu}-\mathcal{L}_mg_{\mu\nu}).
\end{equation}
In Eq. \eqref{Eq:field1}, we denote, in a concise way, $f(R,\mathcal{L}_m)\equiv f$, $\frac{\partial f(R,\mathcal{L}_m)}{\partial R}\equiv f_R$ and $\frac{\partial f(R,\mathcal{L}_m)}{\partial\mathcal{L}_m}\equiv f_{\mathcal{L}_m}$. The energy-momentum tensor considered as,~~$
T_{\mu \nu }=-\frac{2}{\sqrt{-g}}\frac{\delta \left( \sqrt{-g}\mathcal{L}_{m}\right)}{
\delta g^{\mu \nu }}\,$ \cite{LandauLifshitz1975}. To obtain the field equations,  $f(R,\mathcal{L}_{m})=f_{1}(R)+f_{2}(R)G(\mathcal{L}_{m})$ has been considered with $f_{1}(R)$ and $f_{2}(R)$ being arbitrary functions of $R$, $G(\mathcal{L}_{m})$ being a function of matter Lagrangian. \\

In particular, one can retrieve the filed equations of  GR by substituting $f_{1}(R)=\frac{R}{2}$, $f_{2}(R)=1$ and $G(\mathcal{L}_{m})=\mathcal{L}_{m}$ in  Eq. \eqref{Eq:field1}. However, substituting  $f_{2}(R)=1$ and $G(\mathcal{L}_{m})=\mathcal{L}_{m}$, the field equations of $f(R)$ gravity can be retrieved. For a constant $\lambda$, the linear matter-geometry coupling can be considered as, $G(\mathcal{L}_{m})=1+\lambda \mathcal{L}_{m}$. Now, Eq. \eqref{Eq:field1} becomes,
\begin{equation}\label{Eq:field2}
f_{R}R+3\nabla_\mu\nabla^\mu f_{R}-2f=f_{\mathcal{L}_{m}}\left(\frac{1}{2}T-2\mathcal{L}_{m}\right).
\end{equation}
With algebraic manipulations, the term $\nabla_\mu\nabla^\mu f_{R}$ can be eliminated from Eq. \eqref{Eq:field1} and Eq. \eqref{Eq:field2} and hence
\begin{equation}
f_{R}\left(R_{\mu \nu}-\frac{1}{3}R g_{\mu \nu}\right)+\frac{f}{6}-\nabla_{\mu}\nabla_{\nu}f_{R}
=\frac{1}{2}f_{\mathcal{L}_{m}}\left(T_{\mu \nu}-\frac{1}{3}(T-\mathcal{L}_{m}) g_{\mu \nu} \right).
\end{equation}

To construct the cosmological model of the Universe, we consider a flat Friedmann-Lema\^itre-Robertson-Walker (FLRW) space time and the energy momentum tensor that of the perfect fluid. The corresponding expressions are respectively given as,

\begin{equation} \label{flrw}
ds^2=-dt^2+a^{2}(t)\left(dr^{2}+r^2 d\theta^2+r^2\sin^2\theta d\phi^2\right),
\end{equation} 
and 
\begin{align}\label{stressenergy}
T_{\mu \nu}=(\textit{p} + \rho ) u_{ \mu } u_{ \nu }+ \textit{p} g_{ \mu \nu }.
\end{align}

The scaling factor denoted as $a(t)$;~ $u_{\mu}$ are the four velocity vectors along the time directions and satisfy,  $u_{\mu} u_{\nu} g^{\mu \nu }=-1$;~ $\rho $ and $p$ respectively denote the energy density of matter and the isotropic pressure. Now, the field equations of $f(R, \mathcal{L}_m)$ gravity for the FLRW space time and the energy momentum tensor that of perfect fluid  can be obtained as,
\begin{eqnarray}
3H^2f_R+\frac{1}{2}(f-f_R R-f_{\mathcal{L}_m}\mathcal{L}_{m})+3H\dot{f_R}&=&\frac{1}{2}f_{\mathcal{L}_m}\rho, \label{fieldequation1111}\\ 
\dot{H}f_R+3H^2f_R-\ddot{f_R}-3H\dot{f_R}+\frac{1}{2}(f_{\mathcal{L}_m}\mathcal{L}_{m}-f)&=&\frac{1}{2}f_{\mathcal{L}_m}p. \label{fieldequation2}
\end{eqnarray}
Henceforth, an over dot represents the ordinary derivative with respect to time and the Hubble parameter, $H=\frac{\dot{a}}{a}$. The  energy conservation equation for $f(R,\mathcal{L}_m)$ gravity becomes,
\begin{equation}
\dot\rho \left(1+ \frac{2 \rho~f_{\mathcal{L}_m~\mathcal{L}_m}}{f_{\mathcal{L}_m}} \right)+3H(\rho+p)=0. \label{energy conservation}
\end{equation}

In the present analysis, we adopt the standard dust-scaling relation, \(\rho_m(z)=\rho_{m0}(1+z)^3\), to construct a phenomenological description of the background expansion and constrain the model parameters using cosmological observations. This prescription provides a convenient framework for comparing their background expansion histories with observational data. However, in the general \(f(R,\mathcal L_m)\) framework, the matter evolution is governed by the modified continuity equation [Eqn. \eqref{energy conservation}], which for pressureless matter with \(\mathcal L_m=\rho_m\) yields

\begin{equation}
\rho_m f_{\mathcal L_m}^{\,2}
=\rho_{m0}f_{\mathcal L_m0}^{\,2}(1+z)^3.
\end{equation}

Consequently, the standard dust-scaling relation is recovered when the matter coupling \(f_{\mathcal L_m}\) remains constant along the evolution, or under other special conditions that lead to the same scaling. We therefore regard the standard scaling adopted here as a phenomenological prescription for constructing and constraining the background expansion history. This approach allows us to examine whether modifications in the matter Lagrangian can provide an effective description of the observed late-time expansion. Accordingly, the results obtained below are interpreted within the adopted background prescription. Investigating the implications of the modified continuity equation for the matter evolution provides a natural extension of the present analysis.

Previously, we mentioned $f(R,\mathcal{L}_{m})=f_{1}(R)+f_{2}(R)G(\mathcal{L}_{m})$ and in one of our previous work\cite{devi2025late} , we have performed dynamical system analysis for the models described below and obtained the accelerating behavior in the stable critical points. The cosmological parameters were expressed in terms of dimensionless variables, and the evolutionary behavior of the models was studied. Motivated with the viability of the two forms of $f(R,\mathcal{L}_{m})$, here, we shall constrain the model parameters and the cosmological parameters using different the cosmological datasets to explore the evolutionary behavior of the Universe and possible easing of $H_0$ tension. \\

\subsection{Model--I:~~The Power-Law Model}
\label{modelI}
We consider the functional form of $f(R,\mathcal{L}_m)$ \cite{Harko_2010, devi2025late} in the form of a power function of matter Lagrangian as,
\begin{equation}
f(R,\mathcal{L}_m)= \frac{R}{2}+\mathcal{L}_m+\alpha \mathcal{L}_m^{b_1}.  \label{form}
\end{equation}
Comparing Eq. \eqref{form} with the general form of $f(R,\mathcal{L}_{m})$, we can observe $f_1(R)=\frac{R}{2}$, $f_2(R)=1$ and $G(\mathcal{L}_m)=\mathcal{L}_m+\alpha \mathcal{L}_m^{b_1}$. The geometric sector is retained in the standard Einstein-Hilbert form $\frac{R}{2}$, so any deviation from GR arises solely from the modified matter sector \cite{Harko_2010,Bhagat_2026_adf2}. Here, $b_1$ be the model parameter and $\alpha$ denotes the coupling parameter that controls the strength of the nonlinear correction to the matter sector. In the limit $\alpha\rightarrow0$, the model naturally reduces to the standard matter sector and recovers the Einstein-Hilbert framework in standard GR. Now, substituting Eq. \eqref{form} in Eq.~\eqref{fieldequation1111} and Eq.~\eqref{fieldequation2}, we obtain 
\begin{eqnarray}
    3H^2(z)&=& 3H_0^2[\Omega_{m0}(1+z)^3+\Omega_{m0}(1+z)^{3}\alpha(2b_1-1)(3H_0^2\Omega_{m0}(1+z)^3]^{b_1}, \label{friedmanneqnfr1}\\
    -3H^2(z)-2\dot{H}(z)&=&-\alpha(b_1-1)[3H_0^2\Omega_{m0}(1+z)^3)]^{b_1} \label{friedmanneqnfr2}.
\end{eqnarray}
For detailed calculations and the expression for $\alpha$, one can refer to \hyperref[Appendix-1]{Appendix-I}. Now, evaluating Eq. \eqref{friedmanneqnfr1} at redshift $z=0$, the Hubble-normalized reduced field equation becomes, 
\begin{equation}\label{E2(z)}
    E(z)=[\Omega_{m0}(1+z)^3+(1-\Omega_{m0})(1+z)^{3b_1}]^{\frac{1}{2}}.
\end{equation}
The deceleration parameter $q(z)$ quantifies the kinematic status of cosmic expansion. For $q>0$, the model corresponds to
deceleration and $q<0$ to acceleration, with transition redshift denoted as $z_{tr}$. So, 
\begin{eqnarray}
    q(z) &=& -1 - \frac{\dot{H}}{H^2} = -1 + \frac{(1+z)H'(z)}{H(z)}, \nonumber\\ 
     &=& \frac{3\Omega_{m0}(1+z)^3 + 3b_1(1-\Omega_{m0})(1+z)^{3b_1}}
    {2\left[\Omega_{m0}(1+z)^3+(1-\Omega_{m0})(1+z)^{3b_1}\right]} - 1, 
    \label{q1}
\end{eqnarray}
where the prime $(')$ denotes the derivative with respect to redshift $(z)$ and $\Omega_{m0}$ be the matter density at present.  In the limit $b_1\rightarrow0$, Eq.~\eqref{q1} reduces identically to the standard $\Lambda$CDM. Now, combining the Friedmann and acceleration equations for a spatially flat FLRW background, the total effective equation-of-state parameter can be written purely in terms of the deceleration parameter as
\begin{eqnarray}\label{EoS}
    \omega_{\rm eff}(z) &=& \frac{2q(z)-1}{3} \nonumber \\
    &=& \frac{\Omega_{m0}(1+z)^3 +b_1(1-\Omega_{m0})(1+z)^{3b_1}}{E^2(z)} - 1
\end{eqnarray}
We note that $\omega_{\rm eff}$ here characterizes the total cosmic fluid matter together with the modified-gravity sector. In the subsequent section, we shall constrain the model parameters $b_1$, $\Omega_{m0}$ using the cosmological data sets. Under the adopted standard dust-scaling prescription, $\rho_m(z)=\rho_{m0}(1+z)^3$, Eqn. \eqref{friedmanneqnfr2} exhibits a background expansion history with the same functional form as that of a cosmological model containing a constant-equation-of-state component. While the model originates from a non-minimal curvature-matter coupling action, its background expansion mapping is algebraically equivalent to a $w\text{CDM}$ model. In particular, the parameter $b_1$ determines the redshift dependence of the additional contribution to the expansion rate and can be associated with an effective equation-of-state parameter \eqref{EoS} $w_{\rm eff}=b_1-1$ for this contribution. The observational constraints on $b_1$ therefore provide insight into the effective behavior of the background expansion: a positive best-fit value of $b_1$ corresponds to a quintessence-like effective component, whereas $b_1=0$ recovers the cosmological-constant-like limit of this parameterization. This interpretation should, however, be understood at the level of the background expansion and does not establish that the full modified-gravity model is dynamically equivalent to a constant-equation-of-state cosmology. Moreover, the standard dust-scaling relation is imposed here as a phenomenological prescription and does not necessarily follow from the modified continuity equation in $f(R,\mathcal{L}_m)$ gravity. The inferred effective behavior should therefore be interpreted within this assumption, while a self-consistent treatment of matter evolution would be needed to assess its implications for the full model.

\subsection*{Model--II:~~The Exponential-Law Model}\label{modelII}
Motivated by our previous reconstruction of a square-root exponential matter coupling and subsequent dynamical-system studies of nonlinear matter interactions, we investigate whether a modified exponential correction can produce a stable late-time accelerating phase without introducing an additional scalar field. For this purpose, we consider the exponential form
$G(\mathcal{L}_m)=\mathcal{L}_m-\beta \mathcal{L}_{m0}e^{-b_2\sqrt{\frac{\mathcal{L}m}{\mathcal{L}{m0}}}}$
\cite{devi2025late}, and hence the $f(R,\mathcal{L}_m)$ can be expressed as \cite{Harko_2010, DEVI2026170597}
\begin{equation}
f(R,\mathcal{L}_m)=\frac{R}{2}+\mathcal{L}_m-\beta \mathcal{L}_{m0}e^{-b_2\sqrt{\frac{\mathcal{L}_m}{\mathcal{L}_{m0}}}},
\end{equation}

where $b_2$, $\beta$ are model parameters and $\mathcal{L}_{m0}$ denotes the present value of the matter Lagrangian. To obtain an observationally tractable background parametrization, we initially impose the standard dust-scaling relation $\rho_m(z)=\rho{m0}(1+z)^3$, where $\rho_{m0}=3H_0^2\Omega_{m0}$. Since the exponential dependence of $f(R,\mathcal{L}m)$ generally makes  density-dependent, this scaling is an assumption rather than a general consequence of the modified continuity equation. For this functional form Eq.~\eqref{fieldequation1111} and Eq.~\eqref{fieldequation2} become
\begin{eqnarray}
    &3H^2(z)&= 3H_0^2[\Omega_{m0}(1+z)^3+\beta \Omega_{m0}[1+b_2(1+z)^{\frac{3}{2}}]e^{-b_2(1+z)^{\frac{3}{3}}}], \label{friedmanneqnf2r1}\\
    &-3H^2(z)&-2\dot{H}(z)=-3H_0^2\Omega_{m0}[(1+\frac{b_2(1+z)^{\frac{3}{2}}}{2})\beta e^{-b_2(1+z)^{\frac{3}{2}}}] \label{friedmanneqnf2r2}.
\end{eqnarray}
For detailed calculations and the expression for $\beta$, one can refer \hyperref[Appendix-II]{Appendix-II}. By evaluating Eq. \eqref{friedmanneqnf2r1} at redshift $z=0$, we reduce the equation as the Hubble-normalized field equation as,
\begin{equation}\label{E2exp}
    E(z)=\left[\Omega_{m0}(1+z)^3+(1-\Omega_{m0})[\frac{(1+b_2(1+z)^{\frac{3}{2}})e^{-b_2(1+z)^{\frac{3}{2}}}}{(1+b_2)e^{-b_2}}]\right]^{\frac{1}{2}}.
\end{equation}
Using the Hubble-normalized field equation [Eq.\eqref{E2exp}], the corresponding deceleration parameter becomes,
\begin{equation}
    q(z) = \frac{3(1+z)^3}{2E^2(z)}
    \left[\Omega_{m0} - \frac{b_2^2(1-\Omega_{m0})\,
    e^{-b_2(1+z)^{3/2}}}{2(1+b_2)\,e^{-b_2}}\right] - 1 ,
    \label{q2simp}
\end{equation}
As $b_2\rightarrow0$ the
bracketed correction term vanishes and Eq.~\eqref{q2simp} reduces
identically to the standard $\Lambda$CDM deceleration parameter. Using Eq.\eqref{q2simp}, the effective equation-of-state parameter can be obtained as,
\begin{equation}
\omega_{\,\rm eff}(z) = \frac{(1+z)^3}{E^2(z)}
    \left[\Omega_{m0} - \frac{b_2^2(1-\Omega_{m0})\,
    e^{-b_2(1+z)^{3/2}}}{2(1+b_2)\,e^{-b_2}}\right] - 1 .
    \label{w2simp}
\end{equation}

The asymptotic behavior of the exponential-law model provides key insights into the cosmological dynamics of the non-linear matter coupling. In the limit $b_2 \to 0$, Eq.~\eqref{E2exp} smoothly reduces to $E^2(z) = \Omega_{m0}(1+z)^3 + (1 - \Omega_{m0})$, exactly recovering the standard $\Lambda$CDM expansion history. In the far-future limit ($z \to -1$), the factor $b_2(1+z)^{3/2} \to 0$, causing the numerator $(1 + b_2(1+z)^{3/2})e^{-b_2(1+z)^{3/2}}$ to approach unity. Consequently, $E(z)$ approaches a finite non-zero constant, which corresponds to a true late-time de Sitter accelerating attractor with $q(z \to -1) \to -1$ and $\omega_{\rm eff}(z \to -1) \to -1$. At high redshifts ($z \gg 1$), matter scaling dominates over the coupling term, naturally restoring early-time matter domination ($q \to 1/2$, $\omega_{\rm eff} \to 0$). At the present epoch ($z = 0$), the model yields $q_0 \approx -0.55$, providing a smooth transition from early matter domination to late-time cosmic acceleration.

\section{Observational Data And Methodology}\label{III}
In this section, we describe briefly, the cosmological datasets to be used and the techniques involved in it. First, we use Cobaya (COde for BAYesian Analysis) \cite{Torrado_2021}, which provides specialized sampler algorithms that include: evaluation for single-point pipeline validation, minimize (e.g. BOBYQA \cite{cartis2018improvingflexibilityrobustnessmodelbased, Cartis_2021}) for locating minimum $\chi^2$ configurations, $MCMC$ for posterior distribution sampling, and Polychord \cite{Handley2015, Handley_2015} for nested sampling and Bayesian evidence calculations. For the primary parameter constraints, we employed the $MCMC$ sampler with an adaptive proposal covariance matrix following the methodology established in CosmoMC \cite{Lewis2002, Lewis_2013, neal2005takingbiggermetropolissteps}, running chains until achieving convergence at $R - 1 < 0.01$. Also, to perform a rigorous statistical model comparison beyond information criteria, nested samplers like PolyChord has been employed to compute fully integrated Bayesian evidence ($\ln \mathcal{Z}$) and evaluate the Bayes factor ($\Delta \ln \mathcal{Z}$) against the baseline model.

\begin{itemize}
\item\textbf{Cosmic Chronometers (CC):} It consists of 31 model-independent $H(z)$ measurements obtained from differential ages of passively evolving early-type galaxies~\cite{Zhang2014, Moresco_2016, Simon_2005, M_Moresco_2012,Jimenez_2002,Moresco_2016}, spanning the redshift range $0.07\lesssim z\lesssim 1.97$. The sample size $N=31$ enters the likelihood calculation independently; for model selection under CC alone, the small sample corrected Akaike Information Criterion ($AIC_c$) is reported to prevent under-penalization due to $N=31$. The corresponding likelihood becomes,
\begin{equation}
    \chi^2_{\rm CC} = \sum_{i=1}^{31}
    \frac{\left[H_{\rm th}(z_i,\boldsymbol{\theta}) - H_{\rm obs}(z_i)\right]^2}
    {\sigma_{H}^2(z_i)} ,
    \label{chi2CC}
\end{equation}
where $H_{\rm th}(z_i,\boldsymbol{\theta})$ is the model-predicted Hubble parameter and $\sigma_{H}(z_i)$ are the observational uncertainties.

\item \textbf{Type Ia Supernovae($SN^+$):} The SN$^+$ compilation of 1701 light curves obtained from 1550 distinct Type Ia supernovae (SN$^{+}$)~\cite{Scolnic_2018, Scolnic_2022, Brout_2022}, with the redshift range $0.001<z<2.3$. The theoretical distance modulus is the difference between the apparent magnitude $m$ and its absolute magnitude $ M$. So,
\begin{equation}
    \mu(z_i,\boldsymbol{\theta}) = m-M= 5\log_{10}\left[{D_L(z_i,\boldsymbol{\theta})}\right] + 25 ,
\end{equation}
where $D_L(z_i,\boldsymbol{\theta}) = (1+z_i)\,c\!\int_0^z \frac{dz'}{H(z',\boldsymbol{\theta})}$. $D_L$ is the luminosity distance of the supernovae. The corresponding $\chi^2$ becomes
\begin{equation}
    \chi^2_{\rm SN} = \Delta\mu(z_i,\boldsymbol{\theta})^T\, \mathbf{C}_{\rm SN}^{-1}\,
    \Delta\mu(z_i,\boldsymbol{\theta}),
    \label{chi2SN}
\end{equation}
where $\Delta\mu_i(z_i,\boldsymbol{\theta}) = \mu(z_i,\boldsymbol{\theta})  - \mu_{obs}(z_i)$. $\mathbf{C}_{\rm SN}$ is the full systematic-plus-statistical
covariance matrix. The supernovae absolute magnitude $M$ is treated as a nuisance parameter and is analytically marginalized out over the covariance matrix $\mathbf{C}_{\rm SN}$ within the likelihood routine following standard marginalization protocols. Consequently, $M$ does not enter as an additional sampled MCMC degree of freedom, keeping the baseline parameter counts strictly at $k=2$ for flat $\Lambda\text{CDM}$ ($H_0, \Omega_{m0}$) and $k=3$ for the extended models ($H_0, \Omega_{m0}, b_1 / b_2$). 

\item \textbf{Baryon Acoustic Oscillations(BAO)(DESI-DR2):} We incorporate the latest high-precision BAO measurements from the Dark Energy Spectroscopic Instrument Data Release 2 (DESI-DR2) \cite{DESI:2025dr2}. The DESI-DR2 dataset provides spatial standard-ruler measurements spanning the redshift range $0.295 < z < 2.33$, mapping cosmic expansion across Bright Galaxy Samples (BGS), Luminous Red Galaxies (LRGs), Emission Line Galaxies (ELGs), Quasars (QSOs) \cite{Abdul_Karim_2025}, and the Lyman-$\alpha$ forest \cite{DESI:2025lya}. By evaluating the ratio of the comoving transverse distance $D_M(z)/r_{\rm drag}$, the line-of-sight Hubble distance $D_H(z) = c/H(z)$, and the spherical volume-averaged distance ratio $D_V(z)/r_{\rm drag}$, relative to the sound horizon at the drag epoch $r_{\rm drag}$, we perform $MCMC$ sampling to construct the posterior distributions. The $r_{\rm drag}$ is evaluated dynamically as a derived parameter within our custom theory code by integrating the photon-baryon sound speed $c_s(a)$:
\begin{equation}
r_{\rm drag} = c \int_{0}^{a_{\rm drag}} \frac{c_s(a)}{a^2 H(a)} \, da, \quad c_s(a) = \frac{1}{\sqrt{3 \left(1 + \frac{3\rho_b(a)}{4\rho_\gamma(a)}\right)}}
\end{equation}
where $a_{\rm drag} = 1/(1 + 1059.94)$, $\omega_b = 0.02237$, $\omega_\gamma = 2.4697 \times 10^{-5}$, and $N_{\rm eff} = 3.044$. Because our modified gravity formulations alter expansion exclusively at late times ($z < 2$), early-universe plasma recombination dynamics remain standard, yielding physical sound horizon values of $r_{\rm drag} \approx 145.6 - 147.1\,\text{Mpc}$. These measurements effectively break low-redshift geometric parameter degeneracies and yield stringent bounds on cosmic expansion \cite{lodha2025extendeddarkenergyanalysis, Silva_2025}.

\item \textbf{$H_0$ Prior:}
To assess the impact of a local determination of the Hubble constant on the reconstructed parameters, we additionally perform the analysis with a Gaussian prior on $H_0$ from SH0ES(2022) ~\cite{Riess_2022}  $H_0 = 73.04 \pm 1.04 \ {\rm km\,s^{-1}\,Mpc^{-1}}$, implemented as an additional Gaussian term $\chi^2_{H_0} = (H_0-73.04)^2/1.04^2$ added to the total likelihood. This allows us to compare parameter constraints obtained with a locally anchored  and without an
external local-universe anchor considering a uniform prior, presented respectively as the with locally anchored $H_0$-prior and with uniform $H_0$-prior branches in TABLE-~\ref{tab:aic_bic_h0_power_law} and TABLE~\ref{tab:aic_bic_h0_expo}.

\item \textbf{Combined Likelihood and Sampling:}
Posterior parameter distributions are constructed by evaluating the total log-likelihood function, expressed as the sum of independent dataset contributions:
\begin{equation}
    \ln\mathcal{L}_{\rm tot} = -\tfrac{1}{2}\left(\chi^2_{\rm CC} +
    \chi^2_{\rm SN} + \chi^2_{\rm BAO} \;[+\;\chi^2_{H_0}]\right) ,
    \label{lnLtot}
\end{equation}
where the optional $\chi^2_{\text{$H_0$}}$ term accounts for the local SH0ES calibration constraint in anchored runs \cite{Riess_2022}. Parameter estimation is executed using the Monte Carlo Markov Chain (MCMC) sampler implemented within the Cobaya framework \cite{Torrado_2021}. The individual likelihood components are evaluated with a custom Cobaya likelihood routine that handles the CC Hubble parameter measurements spanning $0.07 \le z \le 1.965$; the built-in sn.pantheonplus module evaluates the Type Ia Supernovae light curves from Pantheon+ over $0.001 \le z \le 2.26$ \cite{Scolnic_2022, Brout_2022}, analytically marginalizing over the supernova absolute magnitude $M_B$; and the bao.desidr2 module evaluates the transverse ($D_M/r_{\rm drag}$) and line-of-sight ($D_H/r_{\rm drag}$) distance ratios from DESI-DR2 \cite{DESI:2025dr2}, querying our background theory module directly for the dynamically computed sound horizon $r_{\rm drag} \approx 145.6 - 147.1\,\text{Mpc}$. For the unanchored regime, uniform uninformative priors are assigned over $H_0 \in [50, 90]\,\mathrm{km\,s^{-1}\,Mpc^{-1}}$, $\Omega_{m0} \in [0.05, 0.6]$, $b_1 \in [-1, 1]$ \hyperref[modelI]{Model-I}, and $b_2 \in [-1, 1]$ \hyperref[modelII]{Model-II}. When the local SH0ES prior is applied, it enters as a direct Gaussian prior $H_0 = 73.04 \pm 1.04\,\mathrm{km\,s^{-1}\,Mpc^{-1}}$ for CC alone, whereas for joint datasets containing Pantheon+, local calibration is incorporated natively through the absolute magnitude prior $M_B$ inside the supernova wrapper. Chain convergence is strictly enforced using the Gelman--Rubin diagnostic, requiring $R - 1 < 0.01$ across four independent chains. Post-processing and parameter contour generation are performed using ChainConsumer, discarding the first $30\%$ of each chain as burn-in. 
\end{itemize}

In order to assess the goodness of fit of the models, we have also calculated the Akaike Information Criterion (AIC), the Bayesian Information Criterion (BIC), and the Deviance Information Criterion (DIC), as well as their comparison with the benchmark $\Lambda\text{CDM}$ model. While the minimum chi-squared ($\chi^2_{\min}$) measures raw goodness-of-fit, information criteria evaluate how effectively candidate models balance fit quality against structural complexity, penalizing unnecessary extra parameters.

\begin{itemize}
    \item \textbf{Akaike Information Criterion (AIC):} It is derived from information theory (Kullback-Leibler divergence), AIC estimates the relative information lost by a given statistical model \cite{Burnham2002}. 
    \begin{equation}
        \text{AIC} = 2k - 2\ln(L_{\max}) \approx 2k + \chi^2_{\min},
    \end{equation}
    where $k$ represents the number of free parameters in the model and $L_{\max}$ is the maximum likelihood value achieved. For small sample sizes satisfying $N/k < 40$, the corrected AIC ($\text{AIC}_c$) is used to prevent overfitting,
    \begin{equation}
        \text{AIC}_c = \text{AIC} + \frac{2k(k+1)}{N - k - 1}.
    \end{equation}

    \item \textbf{Bayesian Information Criterion (BIC):} The BIC serves as a computationally efficient surrogate for the Bayesian evidence by approximating the logarithm of the marginal likelihood. Because its penalty term scales with dataset size as $k\ln(N)$, BIC penalizes additional free parameters more heavily than frequentist criteria when $N$ is large \cite{Schwarz1978},
    \begin{equation}
        \text{BIC} = k\ln(N) - 2\ln(L_{\max}) \approx k\ln(N) + \chi^2_{\min}.
    \end{equation}
    Because $\ln(N) > 2$ for all realistic cosmological sample sizes ($N \ge 8$), BIC places a significantly stronger penalty on model complexity than AIC.

    \item \textbf{Deviance Information Criterion (DIC):} Designed specifically for $MCMC$ Bayesian posterior distributions \cite{Spiegelhalter2002} and accounts for parameter constraints and correlations introduced by priors. 
    \begin{equation}
        \text{DIC} = \bar{D} + p_D = 2\bar{D} - D(\hat{\theta}),
    \end{equation}
    where $D(\theta) = -2\ln L(\theta)$ is the deviance, $\bar{D} = \langle D(\theta) \rangle$ is the posterior mean deviance averaged over the $MCMC$ chain, and $D(\hat{\theta})$ is the deviance evaluated at the posterior mean of the parameter vector $\hat{\theta}$, $p_D$ be the effective number of parameters, measuring true model complexity under Bayesian priors.
\end{itemize}

To evaluate the relative performance of the candidate models against the standard $\Lambda\text{CDM}$ benchmark, we compute the relative differences in information criteria $\Delta\text{IC} = \text{IC}_{\text{model}} - \text{IC}_{\Lambda\text{CDM}}, \quad \text{for } \text{IC} \in \{\text{AIC}, \text{BIC}, \text{DIC}\}$. Under this convention, a negative difference ($\Delta\text{IC} < 0$) indicates that the candidate model is statistically favored over $\Lambda\text{CDM}$, while a positive value ($\Delta\text{IC} > 0$) indicates a preference for the baseline scenario. We interpret these relative differences using established statistical thresholds: for AIC and DIC, $0 \le |\Delta\text{IC}| \le 2$ indicates substantial support essentially equivalent performance, $4 \le |\Delta\text{IC}| \le 7$ represents considerably less support for the model with the higher score, and $|\Delta\text{IC}| > 10$ indicates virtually no observational support \cite{Burnham2002}. Similarly, for BIC, $|\Delta\text{BIC}| \le 2$ denotes weak or negligible evidence, $2 < |\Delta\text{BIC}| \le 6$ indicates positive support, $6 < |\Delta\text{BIC}| \le 10$ signifies strong evidence, and $|\Delta\text{BIC}| > 10$ provides decisive evidence in favor of the lower-scoring model \cite{KassRaftery1995}.

\section{Parameter Estimation}\label{IV}
To constrain and analyze the parameters of the cosmological models mentioned in Section-\ref{II}, we shall use Cobaya \cite{Torrado_2021} as a cosmological likelihood wrapper. This framework enabled us to find intermediate calculations and exploits the different computational speeds of the model components during sampling. We perform joint likelihood constraints using combinations of CC, ($\text{SN}^+$), and BAO from DESI-DR2. To examine the impact of local measurements on the posterior parameter distributions and cosmic tensions, we execute the sampling process under two distinct configurations: (a) without a locally anchored $H_0$ prior allowing an unconstrained local expansion rate, and (b) with a locally measured gaussian $H_0$  prior incorporating the Hubble constant measurement \cite{Riess_2019} ($H_0 = 73.04 \pm 1.04 \text{ km s}^{-1} \text{ Mpc}^{-1}$).\\

\begin{figure*}[t!]
    \centering
    \begin{subfigure}[t]{0.48\textwidth}
        \centering
        \includegraphics[width=\textwidth]{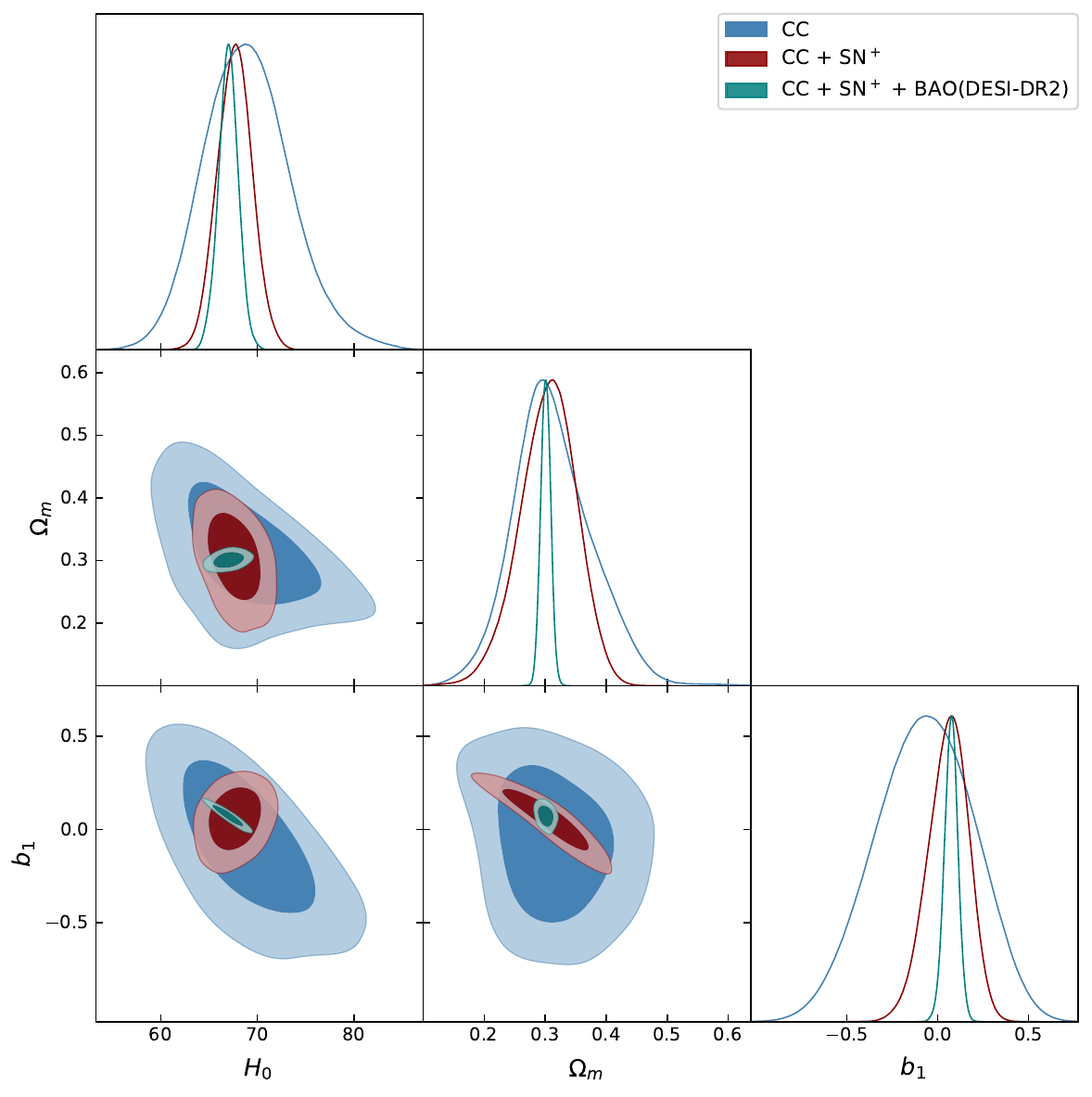}
        \caption{With uniform $H_0$ prior}
        \label{fig:power_law}
    \end{subfigure}
    \hfill
    \begin{subfigure}[t]{0.48\textwidth}
        \centering
        \includegraphics[width=\textwidth]{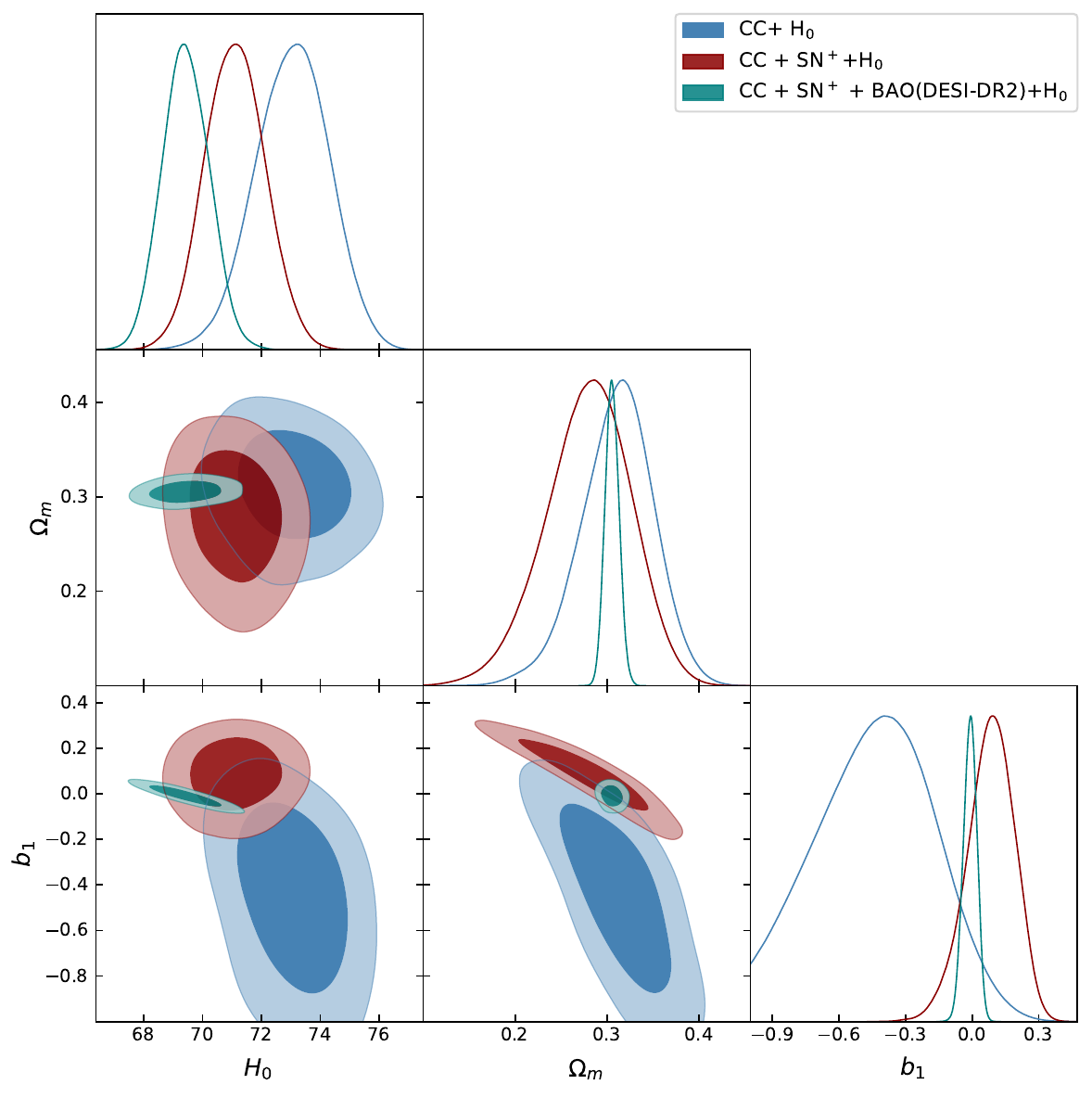}
        \caption{With Gaussian $H_0$ prior from SH0ES(2022)}
        \label{fig:power_law_H0}
    \end{subfigure}
    \caption{Posterior constraints from CC + SN$^+$ + BAO(DESI-DR2) for \hyperref[modelI]{Model-I}.}
    \label{fig:combined_triangle_power}
\end{figure*}

For \hyperref[modelI]{Model-I}, we shall constrain the model parameters $H_0$, $\Omega_m$, $b_1$ those appear in the power law form of $f(R, \mathcal{L}_m)$ model. The marginal posterior distributions of 1-D and 2-D for the parameters are given in FIG.~\ref{fig:combined_triangle_power}. The constrained cosmological parameters, mean values, $1\sigma$ uncertainties, minimum chi-squared ($\chi^2_{\min}$), $\text{AIC}$, and $\text{BIC}$ are summarized in TABLE~\ref{tab:aic_bic_h0_power_law}. Under the unanchored late-time expansion dataset ($\text{CC} + \text{SN}^+ + \text{BAO(DESI DR-2)}$ with a uniform $H_0$ prior, the constrain of the Hubble parameter becomes $H_0 = 66.998 \pm 1.037\,\mathrm{km\,s^{-1}\,Mpc^{-1}}$, the present-day matter density is $\Omega_m = 0.3008 \pm 0.0081$. The power-law modification parameter, $b_1 = 0.0732 \pm 0.0373$, exhibits a minor $\sim 1.96\sigma$ deviation from zero, remaining broadly consistent with standard $\Lambda\text{CDM}$ expansion dynamics. \\

\begin{table}[htbp]
\tiny
\centering
\caption{Constrained values of the parameters with and without a locally anchored $H_0$ prior along with AIC, BIC and DIC for \hyperref[modelI]{Model-I}.}
\label{tab:aic_bic_h0_power_law}
\begin{tabular}{lcccccccccc}
\hline
Dataset & $H_0$ & $\Omega_m$ & $b_1$ & $\chi^2_{\min}$ & AIC & BIC & DIC & $\Delta$AIC & $\Delta$BIC & $\Delta$DIC  \\
\hline
\multicolumn{7}{c}{With uniform $H_0$ prior} \\
\hline
CC & 68.97420 $\pm$ 4.48790   & 0.31304 $\pm$ 0.06526 & -0.07113 $\pm$ 0.26497 & 14.5801 & 21.4689 & 24.8820 & 19.8729 & 2.516 & 3.49& 1.466 \\
CC + SN$^+$  & 67.65546 $\pm$ 1.79035 & 0.30444 $\pm$ 0.04659 & 0.05605 $\pm$ 0.11105 & 1417.5194 & 1423.5194 & 1439.8905 & 1423.3871 & -5.039 & 0.419 & -5.192\\
CC + SN$^+$ + BAO & 66.99781 $\pm$ 1.03668 & 0.30075 $\pm$ 0.00810 & 0.07321 $\pm$ 0.03732 & 1426.8376 & 1432.8376 & 1449.2225 & 1432.9353 & -12.066 & -6.604 & -12.169 \\
\hline
\multicolumn{7}{c}{With a locally anchored gaussian $H_0$ prior from SH0ES(2022)} \\
\hline
CC & 73.05577 $\pm$ 1.24470 & 0.31097 $\pm$ 0.03948 & -0.43170 $\pm$ 0.25591 & 14.4797 & 21.3369 & 24.8770 & 20.0538 & -0.242 & 0.78 & -1.282 \\
CC + SN$^+$ & 71.10825 $\pm$ 1.01707 & 0.27944 $\pm$ 0.04613 & 0.08224 $\pm$ 0.10698 & 1423.8384 & 1429.8384 & 1446.2112 & 1429.7625 & 1.281 & 6.739 & 1.373 \\
CC + SN$^+$ + BAO & 69.43916 $\pm$ 0.79741 & 0.30537 $\pm$ 0.00758 & -0.00942 $\pm$ 0.03026 & 1440.8593 & 1446.8593 & 1463.2459 & 1446.6467 & 1.954 & 7.416 & 1.672 \\
\hline
\end{tabular}
\end{table}

\begin{figure*}[t!]
    \centering
    \begin{subfigure}[t]{0.48\textwidth}
        \centering
        \includegraphics[width=\linewidth]{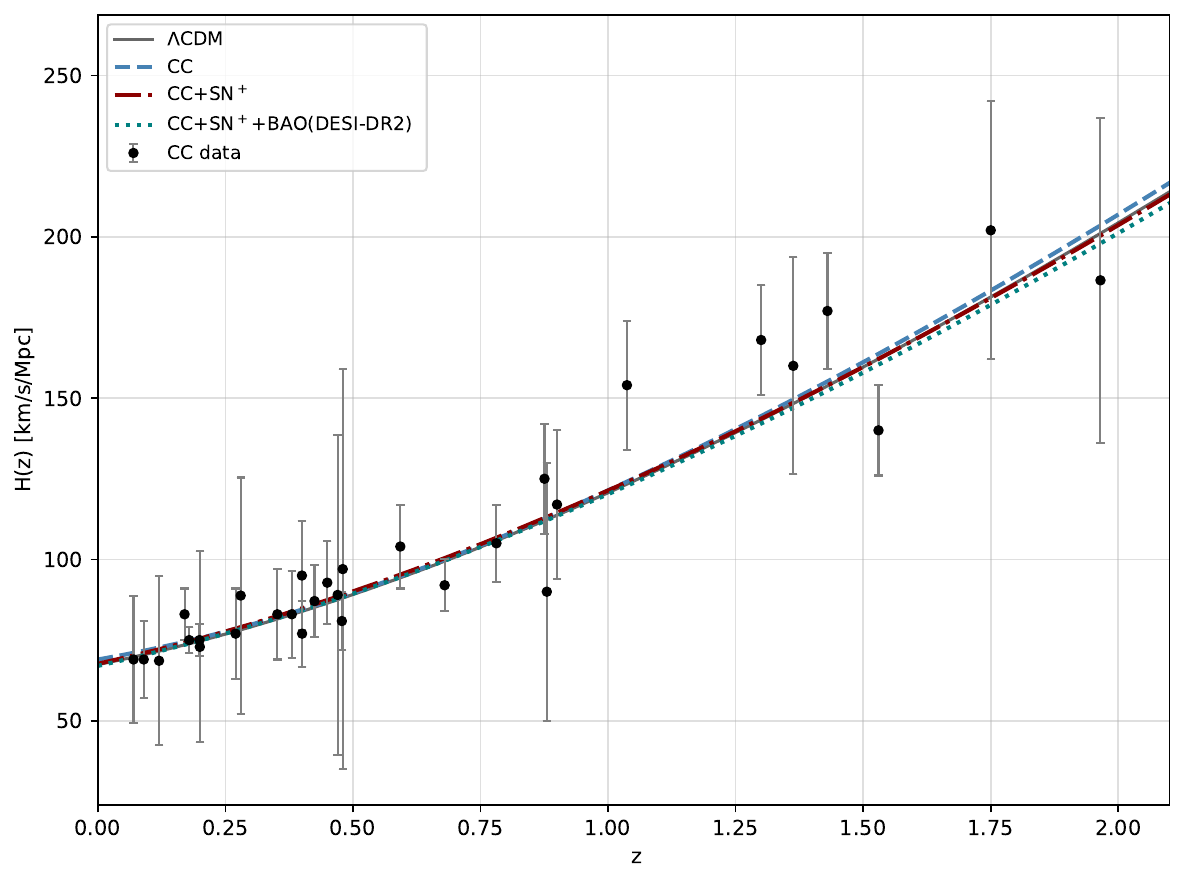}
        \caption{With uniform $H_0$ prior}
        \label{fig:Hz_power_law}
    \end{subfigure}
    \hspace{0.05cm}
    \hfill
    \begin{subfigure}[t]{0.48\textwidth}
        \centering
        \includegraphics[width=\linewidth]{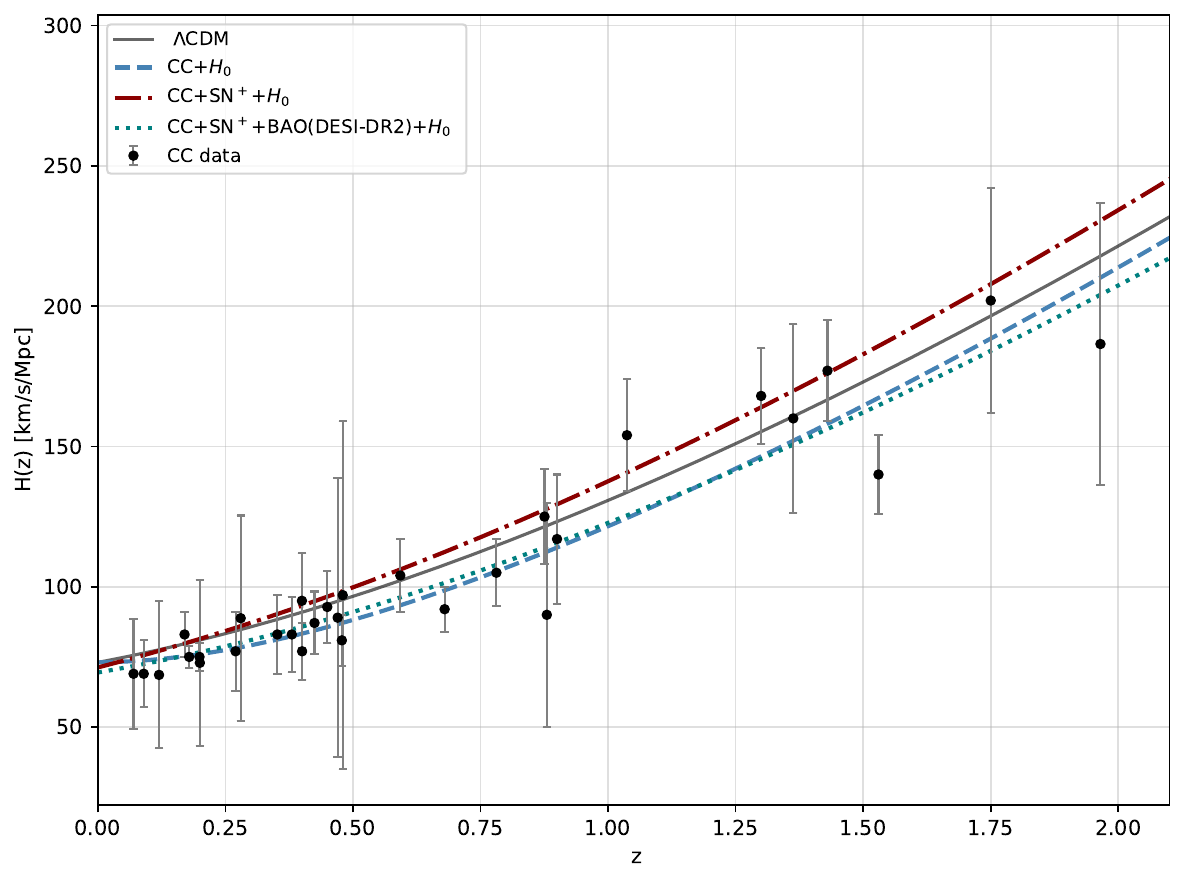}
        \caption{With Gaussian $H_0$ prior from SH0ES(2022) }
        \label{fig:Hz_power_law_H0}
    \end{subfigure}
    \caption{Hubble Parameter vs redshift  \hyperref[modelI]{Model-I} with constrained parametric values}.
    \label{fig:combined_H(z)vsz_power-law}
\end{figure*}

As illustrated in the triangle plots, there is a pronounced parameter degeneracy between $H_0$ and $\Omega_m$, as well as between $H_0$ and $b_1$. The incorporation of BAO(DESI-DR2) observations plays a critical role in breaking these degeneracy directions, significantly shrinking the joint confidence contours $1\sigma$ and $2\sigma$  compared to the individual subsets of $\text{CC}$ and $\text{CC} + \text{SN}^+$ . In contract, when the expansion scales with the local Cepheid prior ($H_0 = 73.04 \pm 1.04\,\mathrm{km\,s^{-1}\,Mpc^{-1}}$), a clear dataset tension emerges after incorporating BAO(DESI DR-2) data. The BAO baseline strongly pulls the preferred expansion rate towards lower values ($H_0 = 69.439 \pm 0.797\,\mathrm{km\,s^{-1}\,Mpc^{-1}}$), while keeping the matter density parameter stable at $\Omega_m = 0.30537 \pm 0.00758$ and driving $b_1$ back to $-0.0094 \pm 0.0303$, which is consistent with zero within $0.3\sigma$. This orthogonal shift between posterior contours $\text{CC}$ and the combined dataset in parameter space underscores the intrinsic conflict between local distance ladder calibrations and early-universe sound horizon measurements mapped by BAO at intermediate redshifts. The coupling parameter $b_1$ shows a slight positive value ($0.07321 \pm 0.03732$) under the unconstrained baseline. However, introducing the local $H_0$ measurement pulls $b_1$ slightly negative $-0.00942 \pm 0.03026$, remaining fully consistent with standard GR within $1\sigma$ credible region. Also, FIG. ~\ref{fig:combined_H(z)vsz_power-law} shows the Hubble evolution $H(z)$ against observational datasets over $0 \le z \le 2.5$, demonstrating that the prior-assisted fit effectively elevates the $z=0$ intercept to accommodate local $H_0$ observations.\\

To assess the physical necessity of an additional degree of freedom $b_1$, we performed a rigorous statistical model comparison against the baseline flat $\Lambda\text{CDM}$ model using the $\Delta\text{AIC}$, $\Delta\text{BIC}$, and $\Delta\text{DIC}$ criteria. For the unanchored combined dataset ($\text{CC} + \text{SN}^+ + \text{BAO(DESI DR-2)}$), the power-law model yields $\Delta\text{AIC} = -12.066$, $\Delta\text{BIC} = -6.604$, and $\Delta\text{DIC} = -12.169$. According to the Kass-Raftery and Jeffreys interpretation scales \cite{Kass01061995, Jeffreys:1939xee}, a value of $\Delta\text{BIC} = -6.604$ constitutes strong statistical evidence favoring the power-law model over $\Lambda\text{CDM}$. Here, the significant fit improvement ($\Delta\chi^2_{\min} \approx -14.07$) easily overcomes the single-parameter sample-size penalty ($\Delta k \cdot \ln(N) \approx +7.46$) levied by the BIC metric. Conversely, when the local SH0ES(2020) $H_0$ prior is imposed alongside BAO(DESI-DR2), the strict acoustic sound horizon scale $r_{\rm drag}$ forces the modification parameter to collapse to $b_1 = -0.0094 \pm 0.0303 \approx 0$. Because \hyperref[modelI]{Model-I} degenerates back to the standard $\Lambda\text{CDM}$ expansion history ($\Delta\chi^2_{\min} \approx 0$), it incurs information criteria penalties of $\Delta\text{AIC} = +1.954$, $\Delta\text{BIC} = +7.416$, and $\Delta\text{DIC} = +1.672$. A value of $\Delta\text{BIC} = +7.416$ ($> 6$) provides strong statistical evidence in favor of baseline $\Lambda\text{CDM}$. This transition in $\Delta\text{BIC}$ demonstrates that because the power-law model modifies gravity exclusively at late times ($z < 2$), it lacks the early-universe flexibility needed to simultaneously reconcile the SH0ES $H_0$ calibration with DESI BAO angular distances without forcing $b_1 \to 0$ and incurring an unused parameter penalty.  This significant increase in $\Delta\text{BIC}$ demonstrates that the power-law parametrization lacks the necessary dynamical flexibility to simultaneously reconcile the SH0ES $H_0$ calibration with DESI BAO angular diameter distances without incurring a severe likelihood penalty. Consequently, under the anchored regime, Occam's razor \cite{Liddle_2007} strongly penalizes extended parametrization, confirming that standard $\Lambda\text{CDM}$ remains the most statistically parsimonious model.\\

To visually inspect the dynamical evolution of the late-time power-law model against observational background data, we present the reconstructed Hubble parameter trajectories $H(z)$ alongside the $31$ CC measurements in FIG. ~\ref{fig:Hz_power_law} and  FIG. \ref{fig:Hz_power_law_H0}. Under a uniform $H_0$ prior FIG.~\ref{fig:Hz_power_law}, the theoretical curves for $\Lambda\text{CDM}$, $\text{CC}$, $\text{CC}+\text{SN}^+$, and the full joint dataset $\text{CC}+\text{SN}^++\text{BAO (DESI DR-2)}$ are virtually indistinguishable across $z \in [0, 2.5]$, passing smoothly through the CC measurements and converging to an unanchored Hubble constant of $H_0 = 66.998 \pm 1.037\,\mathrm{km\,s^{-1}\,Mpc^{-1}}$. In contrast, when the local SH0ES  $H_0$ prior is imposed (FIG.~\ref{fig:Hz_power_law_H0}), a marked trajectory dispersion emerges: the $\text{CC}+H_0$ fit shifts upward to $H_0 = 73.06 \pm 1.24\,\mathrm{km\,s^{-1}\,Mpc^{-1}}$, driving $b_1 = -0.432 \pm 0.256$ to negative values to match high-redshift trends, whereas adding DESI BAO ($\text{CC}+\text{SN}^++\text{BAO(DESI DR-2)}+H_0$) firmly pulls the trajectory back to $H_0 = 69.44 \pm 0.80\,\mathrm{km\,s^{-1}\,Mpc^{-1}}$ and $b_1 = -0.0094 \pm 0.0303$. This visual divergence highlights the structural inability of the power-law modification to resolve the Hubble tension without incurring severe likelihood penalties. Nevertheless, the power-law model demonstrates notable physical viability in the unanchored regime: for the joint $\text{CC}+\text{SN}^++\text{BAO(DESI DR-2)}$ dataset, it achieves a lower minimum chi-squared $\chi^2_{\min} = 1426.8376$ than $\Lambda\text{CDM}$ ($\Delta\chi^2_{\min} \approx -14.07$) [TABLE~ \ref{tab:aic_bic_h0_power_law}, TABLE~\ref{tab:aic_bic}]. It accommodates standard cosmology as a limiting case $b_1 = 0.0732 \pm 0.0373$ within $\sim 1.96\sigma$, yields matter density estimates $\Omega_m = 0.3008 \pm 0.0081$ in close agreement with canonical values, and provides a smooth, non-singular expansion trajectory free of high-redshift instabilities.\\

\begin{figure*}[t!]
    \centering
    \begin{subfigure}[t]{0.48\textwidth}
        \centering
        \includegraphics[width=\textwidth]{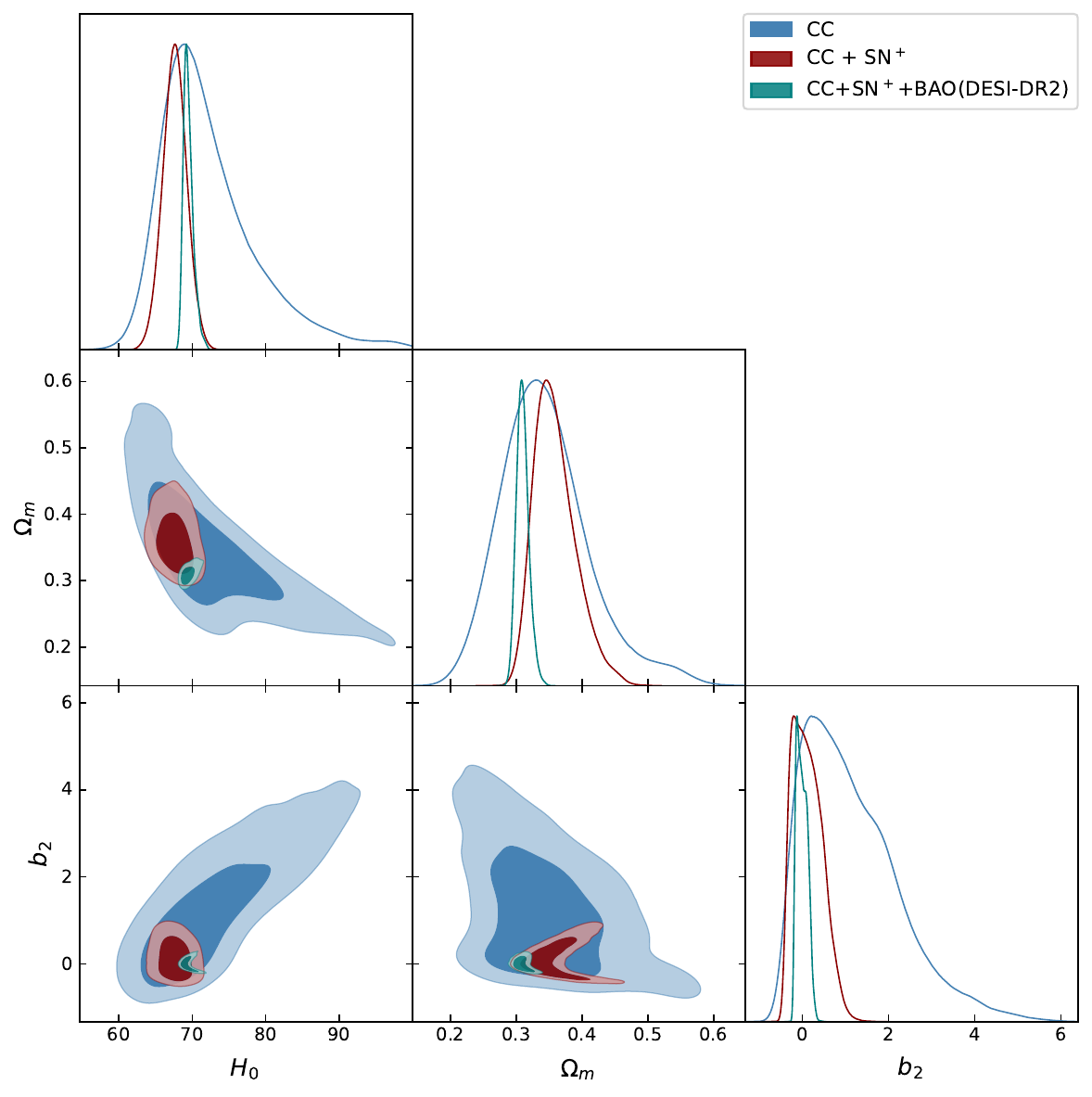}
        \caption{With uniform $H_0$ prior}
        \label{fig:exponential-power}
    \end{subfigure}
    \hfill
    \begin{subfigure}[t]{0.48\textwidth}
        \centering
        \includegraphics[width=\textwidth]{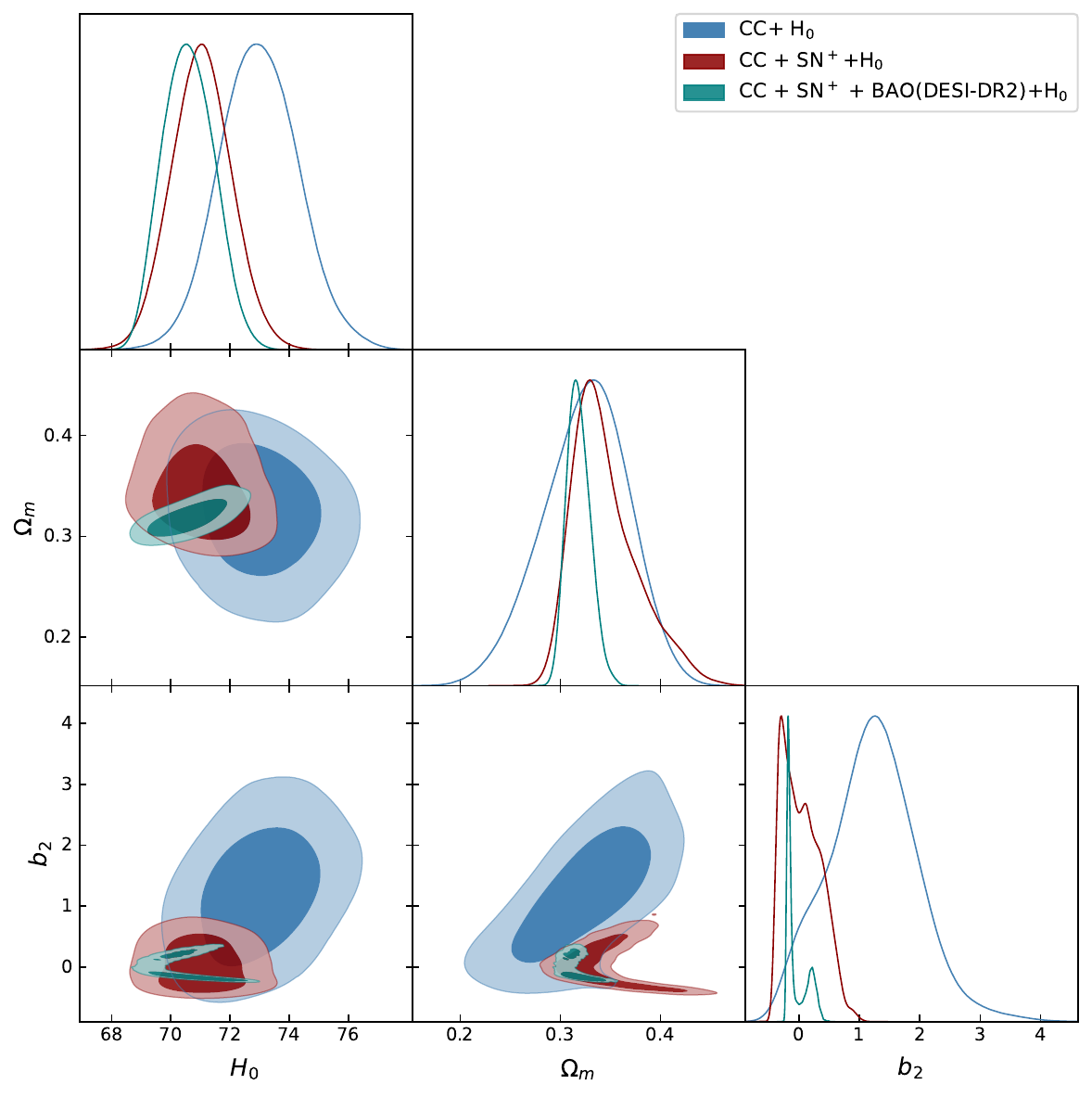}
        \caption{With Gaussian prior from SH0ES(2022) local $H_0$ measurement}
        \label{fig:exponential-power_H0}
    \end{subfigure}
    \caption{Posterior constraints from CC + SN$^+$ + BAO(DESI-DR2) for  model \ref{modelII}. }
    \label{fig:combined_triangle_expo}
\end{figure*}

\vspace{0.2 cm}
For \hyperref[modelII]{Model-II}, the posterior contours for the model parameters $H_0$, $\Omega_m$, $b_2$ are presented with the corresponding marginalized 1-D and 2-D  distributions in FIG.~\ref{fig:combined_triangle_expo} with the statistical breakdown detailed in TABLE.~\ref{tab:aic_bic_h0_expo}. In the unanchored regime assuming a uniform $H_0$ prior, the pure CC dataset weakly constrains the expansion rate to $H_0 = 72.19 \pm 6.71\,\mathrm{km\,s^{-1}\,Mpc^{-1}}$, $\Omega_m = 0.344 \pm 0.068$, and $b_2 = 1.128 \pm 1.117$, exhibiting a broad negative degeneracy in the $H_0$--$\Omega_m$ plane and a positive correlation between $H_0$ and $b_2$. Incorporating Type Ia supernovae ($\text{CC}+\text{SN}^+$) substantially tightens the matter density to $\Omega_m = 0.359 \pm 0.032$, shifts the Hubble constant downward to $H_0 = 67.61 \pm 1.63\,\mathrm{km\,s^{-1}\,Mpc^{-1}}$, and pulls the coupling parameter close to the $\Lambda\text{CDM}$ limit, $b_2 = 0.125 \pm 0.339$. Adding DESI DR-2 BAO data ($\text{CC}+\text{SN}^++\text{BAO}$) breaks the remaining geometric degeneracies, sharply constraining the parameters to $H_0 = 69.46 \pm 0.70\,\mathrm{km\,s^{-1}\,Mpc^{-1}}$, $\Omega_m = 0.309 \pm 0.010$, and $b_2 = -0.004 \pm 0.122$, which recovers the standard $\Lambda\text{CDM}$ baseline ($b_2 = 0$) well within $1\sigma$. When anchoring the expansion scale with the local SH0ES (2022) Gaussian prior, $\text{CC}+H_0$ isolates a strongly coupled regime with $H_0 = 72.99 \pm 1.33\,\mathrm{km\,s^{-1}\,Mpc^{-1}}$, $\Omega_m = 0.325 \pm 0.043$, and $b_2 = 1.214 \pm 0.751$, whereas adding $\text{SN}^+$ and BAO shifts the joint constraints to $H_0 = 71.04 \pm 1.02\,\mathrm{km\,s^{-1}\,Mpc^{-1}}$ ($b_2 = 0.044 \pm 0.314$) for $\text{CC}+\text{SN}^++H_0$ and $H_0 = 70.60 \pm 0.86\,\mathrm{km\,s^{-1}\,Mpc^{-1}}$, $\Omega_m = 0.318 \pm 0.012$, and $b_2 = -0.050 \pm 0.178$ for $\text{CC}+\text{SN}^++\text{BAO}+H_0$. Crucially, the SH0ES-anchored joint contours display a prominent bimodal distribution in $b_2$ straddling $b_2 = 0$ alongside a V-shaped degeneracy in the $\Omega_m$--$b_2$ plane: because of the leading-order exponential correction to the cosmic expansion rate depends quadratically on $b_2^2$ rather than linearly on $b_2$. The background evolution exhibits an approximate reflection symmetry ($b_2 \leftrightarrow -b_2$) around the $\Lambda\text{CDM}$ limit ($b_2 = 0$). When the joint dataset favors a mild departure from $\Lambda\text{CDM}$ to reconcile the high local $H_0$ prior with intermediate-redshift BAO and $\text{SN}^+$ distance ratios, the likelihood surface develops a local saddle barrier at $b_2 = 0$ and splits into two degenerate branches accompanied by a slight upward shift in $\Omega_m$ along both arms. Higher-order cubic corrections break this reflection symmetry, producing a sharper, dominant peak on the negative branch ($-1 < b_2 < 0$), which is phenomenologically favored by late-time observations due to its stronger phantom-like acceleration boost at low redshifts ($z \lesssim 2.5$). However, global physical viability across the entire cosmic history strictly insists on $b_2 \geq 0$, as extrapolating negative $b_2$ values to early times drives the exponential correction factor negative beyond a critical high redshift, violating background positivity.\\ 

\begin{table}[htbp]
\tiny
\centering
\caption{Constrained values of the cosmological and model parameters with and without a locally anchored $H_0$ prior along with AIC, BIC and DIC for \hyperref[modelII]{Model-II}.}
\label{tab:aic_bic_h0_expo}
\begin{tabular}{lcccccccccc}
\hline
Dataset & $H_0$ & $\Omega_m$ & $b_2$ & $\chi^2_{\min}$ & AIC & BIC & DIC & $\Delta$AIC & $\Delta$BIC & $\Delta$DIC \\
\hline
\multicolumn{7}{c}{Uniform $H_0$ prior} \\
\hline
CC & 72.19092 $\pm$ 6.70560  & 0.34395 $\pm$ 0.06847 & 1.12763 $\pm$ 1.11701 & 14.4973 & 21.3862 & 24.7992 & 20.0697 & 2.434 &  3.407 & 1.663\\

CC + SN$^+$  & 67.60844 $\pm$ 1.63345& 0.35871 $\pm$ 0.03221 & 0.12514 $\pm$ 0.33868 & 1417.1787 & 1423.1787 & 1439.5498 & 1422.6856 & -5.111 & 0.078  & -5.893\\

CC + SN$^+$ + BAO & 69.46266 $\pm$ 0.69589 & 0.30935 $\pm$ 0.00952 & -0.00360 $\pm$ 0.12168 & 1430.5355 & 1436.5355 & 1452.9204 & 1435.5506 & -8.368 & -2.907 & -9.554  \\
\hline
\multicolumn{7}{c}{With a locally anchored gaussian $H_0$ prior from SH0ES(2022)} \\
\hline
CC & 72.98879 $\pm$ 1.32662 & 0.32533  $\pm$ 0.04272 & 1.21387  $\pm$ 0.75125 & 14.7669 & 21.6240 & 25.1641 & 21.4523 & -0.398 & 1.067 & 0.116  \\

CC + SN$^+$  & 71.03821 $\pm$ 1.01677 & 0.34481 $\pm$ 0.03332 & 0.04393  $\pm$ 0.31435 & 1423.6046 & 1429.6046 & 1445.9774 & 1429.8634 & 1.047 & 6.505 & 1.474\\

CC + SN$^+$  + BAO & 70.59847 $\pm$ 0.86198 & 0.31823 $\pm$ 0.01203 & -0.04969 $\pm$ 0.17821 & 1434.6112 & 1440.6112 & 1456.9978 & 1443.5037 & -4.294  &  1.168  & -1.471  \\
\hline
\end{tabular}
\end{table}

To evaluate the statistical viability of exponential modification against flat $\Lambda\text{CDM}$, we examine the relative information criteria $\Delta\text{AIC}$, $\Delta\text{BIC}$, and $\Delta\text{DIC}$ detailed in TABLE~\ref{tab:aic_bic_h0_expo}. In the unanchored regime, the standalone CC dataset yields positive values of $\Delta\text{AIC} = 2.43$, $\Delta\text{BIC} = 3.41$, and $\Delta\text{DIC} = 1.66$, indicating that the minor improvement in $\chi^2_{\min}$ does not justify the additional free parameter $b_2$ showing that Occam's razor penalizes the extra free parameter $b_2$ since the marginal reduction in $\chi^2_{\min}$ does not justify the added complexity. Incorporating $\text{SN}^+$ observations substantially improves the fit to $\Delta\text{AIC} = -5.11$ and $\Delta\text{DIC} = -5.89$, indicating positive preference according to the Jeffreys and Kass--Raftery scales, although the stricter logarithmic sample-size penalty imposed by Occam's razor in BIC leaves the comparison statistically neutral at $\Delta\text{BIC} = 0.08$. Combining all unanchored datasets in $\text{CC}+\text{SN}^++\text{BAO}$ overcomes Occam's penalty across all dataset, yielding $\Delta\text{AIC} = -8.37$ and $\Delta\text{DIC} = -9.55$ for strong evidence alongside $\Delta\text{BIC} = -2.91$ for positive evidence on the Kass--Raftery scale in favor of \hyperref[modelII]{Model-II}. Under the SH0ES 2022 $H_0$ prior, the $\text{CC}+H_0$ combination remains statistically indistinguishable from $\Lambda\text{CDM}$ with $\Delta\text{AIC} = -0.40$, $\Delta\text{BIC} = 1.07$, $\Delta\text{DIC} = 0.12$, while $\text{CC}+\text{SN}^++H_0$ incurs strong Bayesian penalization at $\Delta\text{BIC} = 6.51$ on the Kass--Raftery scale in favor of $\Lambda\text{CDM}$ due to tension between the local $H_0$ anchor and intermediate-redshift supernova distances. For the full SH0ES-anchored dataset $\text{CC}+\text{SN}^++\text{BAO}+H_0$, the additional degree of freedom yields positive evidence under AIC at $\Delta\text{AIC} = -4.29$ by partially absorbing the $H_0$ tension however, Occam's $\ln N$ penalty shifts the Bayesian metric into the inconclusive regime favoring $\Lambda\text{CDM}$ at $\Delta\text{BIC} = 1.17$, while the weaker Deviance Information Criterion of $\Delta\text{DIC} = -1.47$ reflects the inflated effective parameter count $p_D$ induced by the bimodal posterior variance of $b_2$.

\begin{figure*}[t!]
    \centering
    \begin{subfigure}[t]{0.48\textwidth}
        \centering
        \includegraphics[width=\linewidth]{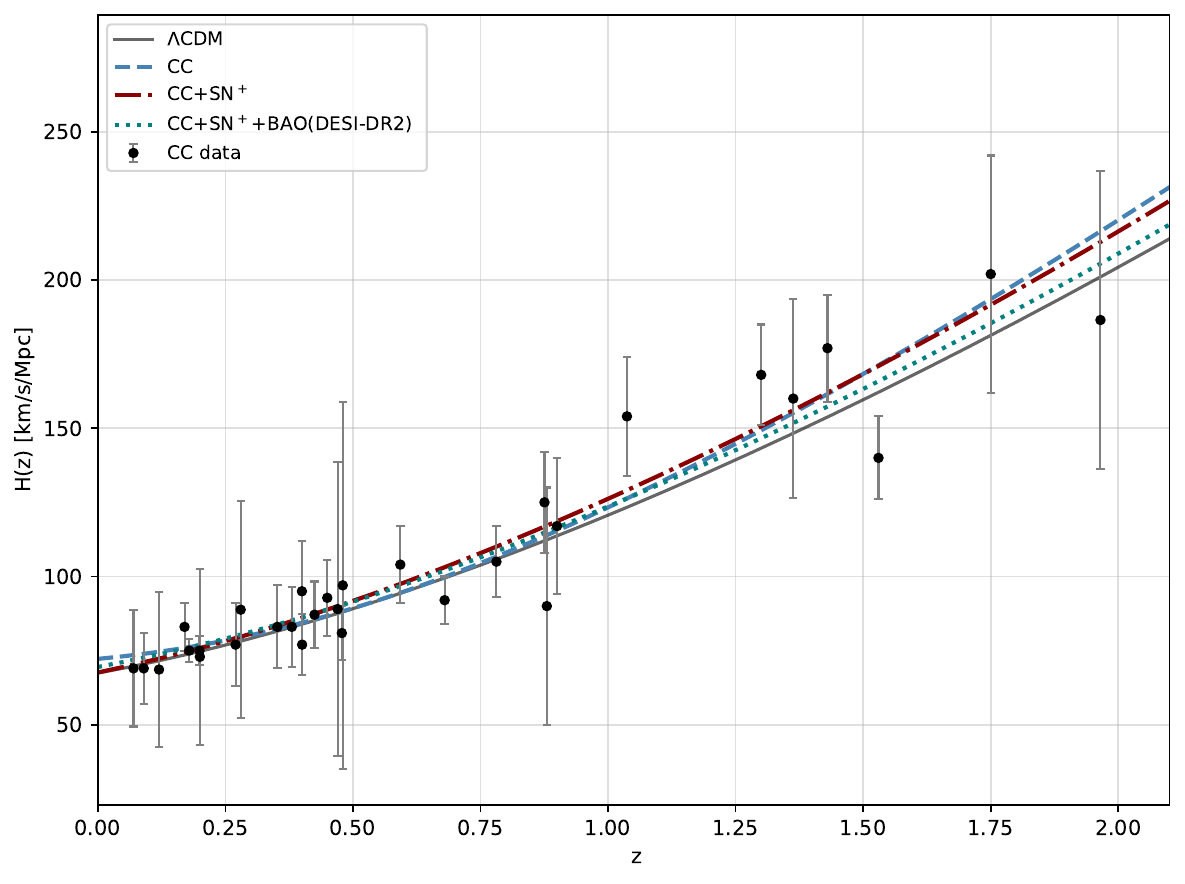}
        \caption{With uniform $H_0$ prior}
        \label{fig:Hz_exponential_model_without_H_0_prior}
    \end{subfigure}
    \hfill
    \begin{subfigure}[t]{0.48\textwidth}
        \centering
        \includegraphics[width=\linewidth]{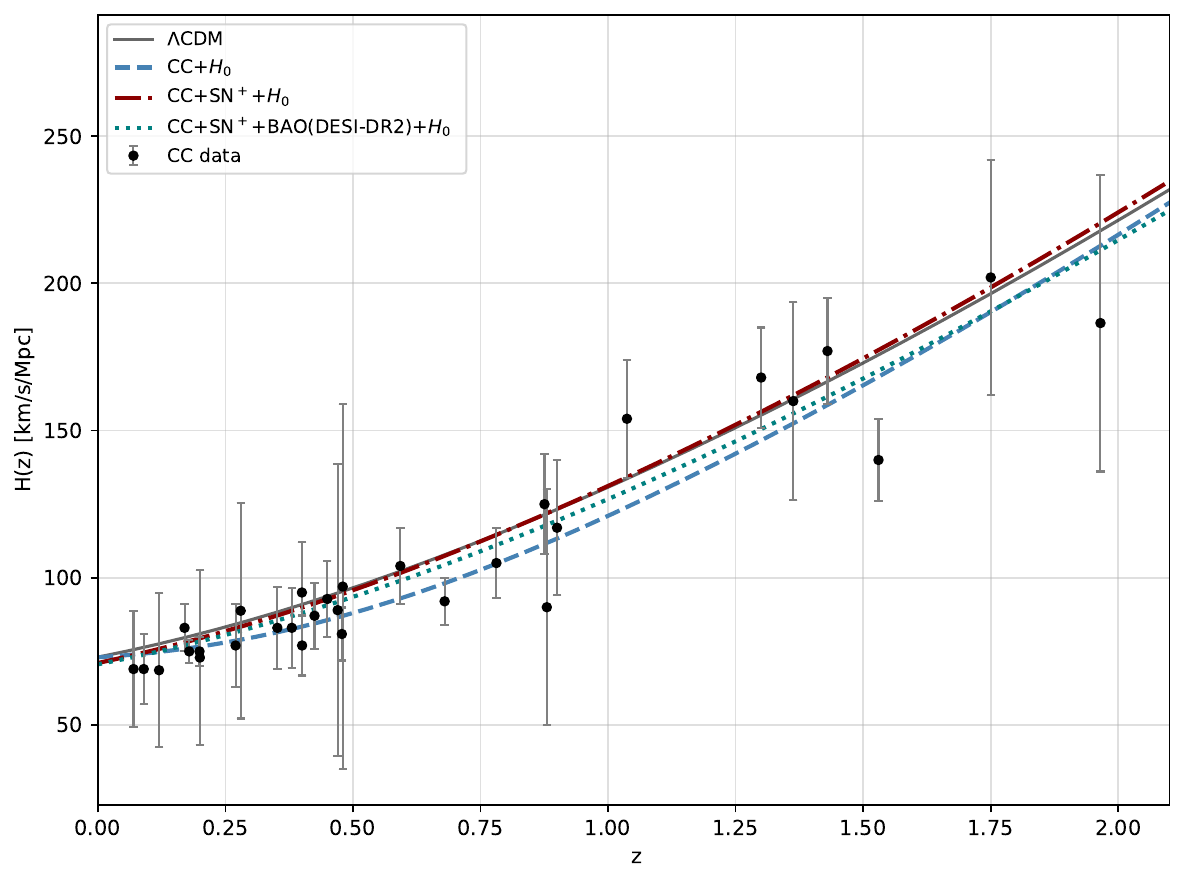}
        \caption{With Gaussian prior from SH0ES(2022) $H_0$ measurement}
        \label{fig:Hz_exponential_model_with_H_0_prior_from_SHOES-2020}
    \end{subfigure}
    \caption{Hubble parameter vs redshift plots for \hyperref[modelII]{Model-II} with the constrained parameteric values.}
    \label{fig:combined_H(z)vsz_exponential-model}
\end{figure*}

As illustrated in FIG.~\ref{fig:combined_H(z)vsz_exponential-model}, the reconstructed Hubble expansion histories $H(z)$ for \hyperref[modelII]{Model-II} exhibit tight clustering across all dataset combinations over the observational range $z \in [0, 2.5]$, closely tracking the cosmic chronometer measurements. In the unanchored regime, the trajectories separate slightly at the present epoch according to their constrained expansion scales, spanning from $H_0 \approx 67.61\,\mathrm{km\,s^{-1}\,Mpc^{-1}}$ for $\text{CC}+\text{SN}^+$ and $H_0 \approx 69.46\,\mathrm{km\,s^{-1}\,Mpc^{-1}}$ for $\text{CC}+\text{SN}^++\text{BAO(DESI-DR2)}$ up to $H_0 \approx 72.19\,\mathrm{km\,s^{-1}\,Mpc^{-1}}$ for CC. At higher redshifts $z > 1.0$, the CC and $\text{CC}+\text{SN}^+$ curves lie systematically above the flat $\Lambda\text{CDM}$ baseline due to their elevated matter densities $\Omega_m \approx 0.344\text{--}0.359$ and positive exponential couplings, whereas the full joint trajectory closely parallels $\Lambda\text{CDM}$ as $b_2$ converges to $-0.004 \pm 0.122$. When the local SH0ES(2022) Gaussian prior is introduced, the $\text{CC}+H_0$ reconstruction accommodates a high present-day rate of $H_0 \approx 72.99\,\mathrm{km\,s^{-1}\,Mpc^{-1}}$ via its large positive coupling $b_2 = 1.214 \pm 0.751$, which flattens the expansion rate and drives the curve below the $\Lambda\text{CDM}$ baseline at intermediate and high redshifts $z \gtrsim 0.5$. Conversely, incorporating $\text{SN}^+$ luminosity distances in $\text{CC}+\text{SN}^++H_0$ pulls the local rate down to $H_0 \approx 71.04\,\mathrm{km\,s^{-1}\,Mpc^{-1}}$ while raising $\Omega_m$ to $0.345$, pushing the high-redshift trajectory back above $\Lambda\text{CDM}$. Including DESI-DR2 BAO constraints in the full joint fit $\text{CC}+\text{SN}^++\text{BAO(DESI-DR2)}+H_0$ mediates this trade-off at $H_0 \approx 70.60\,\mathrm{km\,s^{-1}\,Mpc^{-1}}$ and $b_2 = -0.050 \pm 0.178$, yielding a trajectory that tracks slightly below $\Lambda\text{CDM}$ across $z \gtrsim 0.25$ while partially alleviating the $H_0$ tension with only a mild Bayesian complexity penalty of $\Delta\text{BIC} = 1.17$.

\vspace{0.2cm}

\section{Behavior of Cosmological Parameter}\label{V}
To physically characterize the cosmological dynamics of the reconstructed modified gravity models, in the following, we present the deceleration and EoS parameters for both the models across both unanchored and locally anchored regimes ($H_0$).

\begin{figure*}[t!]
    \centering
    \begin{subfigure}[t]{0.48\textwidth}
        \centering
        \includegraphics[width=\linewidth]{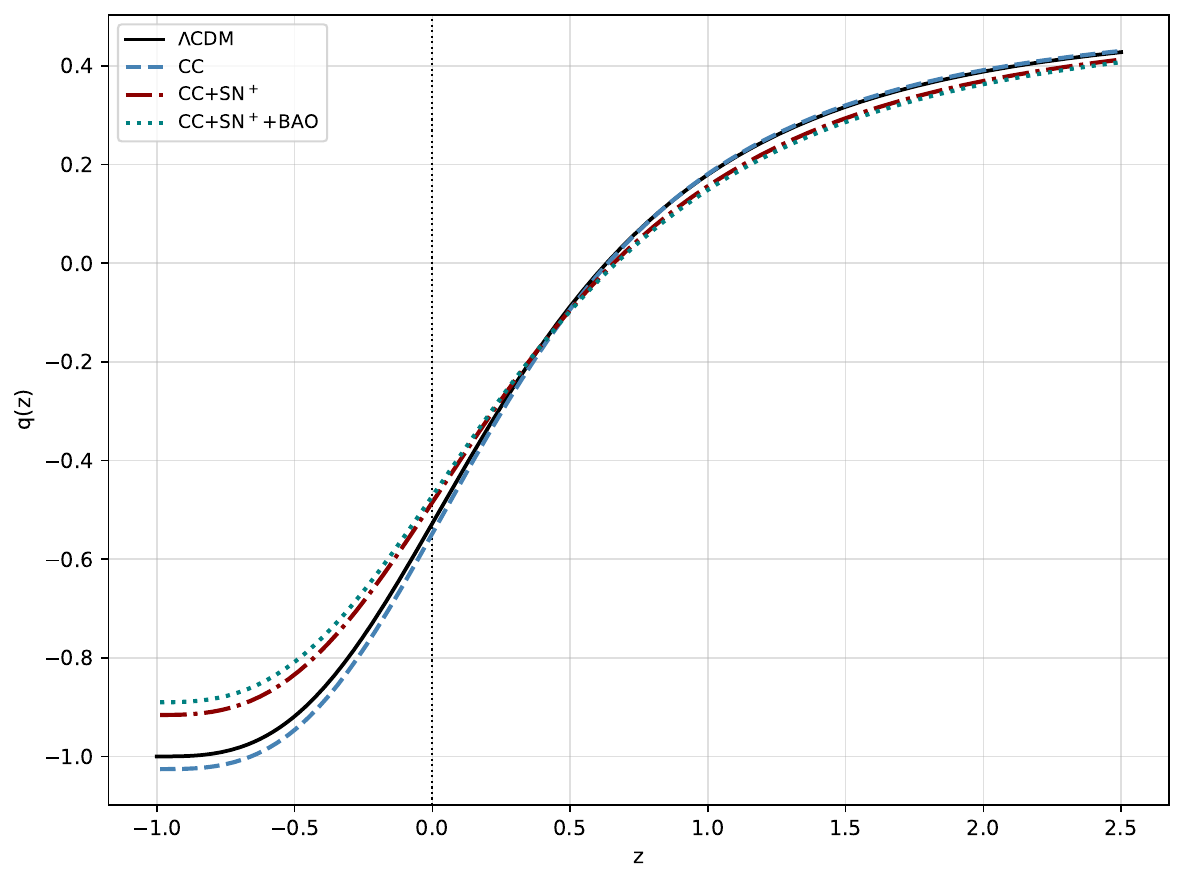}
        \caption{With uniform $H_0$ prior}
        \label{fig:powerlaw model dec without $H_0$ prior}
    \end{subfigure}
    \hspace{0.1 cm}
    \hfill
    \begin{subfigure}[t]{0.48\textwidth}
        \centering
        \includegraphics[width=\linewidth]{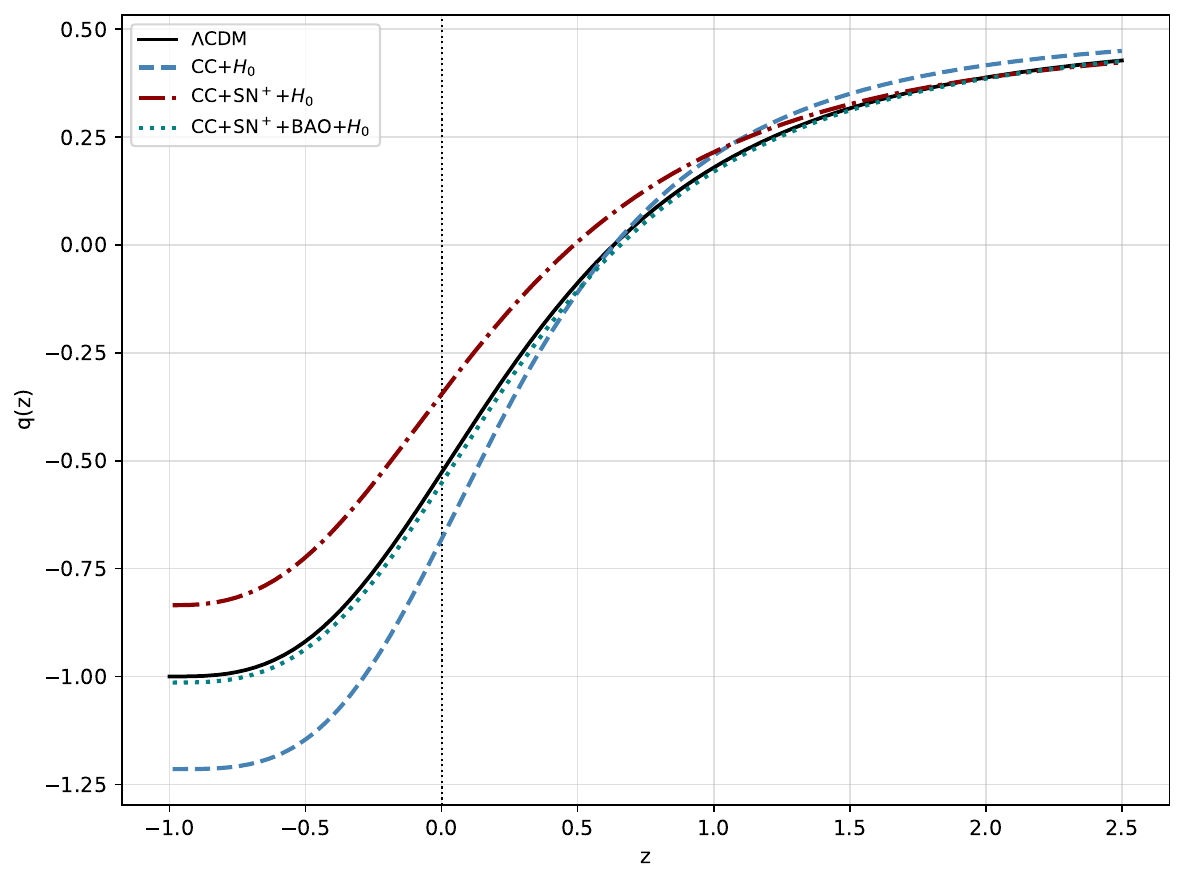}
        \caption{With Gaussian prior from SH0ES(2022) $H_0$ measurement}
        \label{fig:powerlaw model dec with $H_0$ prior from SHOES-2020}
    \end{subfigure}
    \caption{Deceleration parameter vs redshift plots for \hyperref[modelI]{Model-I} with constrained parameter values: FIG.~\ref{fig:powerlaw model dec without $H_0$ prior} is with the uniform $H_0$ prior and FIG.~\ref{fig:powerlaw model dec with $H_0$ prior from SHOES-2020} with a Gaussian $H_0$ prior from SH0ES-2022 measurement}
    \label{fig:Power-law_dec}
\end{figure*}

In the unanchored regime, \hyperref[modelI]{Model-I} yields a smooth, physically consistent transition from an early decelerating, matter-dominated phase with $q(z) > 0$ to a late-time accelerating epoch with $q(z) < 0$, as shown in FIG.~\ref{fig:powerlaw model dec without $H_0$ prior}. Across all unanchored dataset combinations, the deceleration-to-acceleration transition occurs within $z_t \approx 0.62\text{--}0.68$, yielding present-day deceleration values of $q_0 \approx -0.55$ for standalone CC and $q_0 \approx -0.48\,\,\text{to}-0.50$ for the $\text{CC}+\text{SN}^+$ and $\text{CC}+\text{SN}^++\text{BAO}$ combinations. At high redshifts $z > 2.0$, all trajectories asymptotically approach the standard Einstein--de Sitter matter-dominated limit $q \to 0.5$. In the asymptotic future limit $z \to -1$, the CC curve dips slightly below the de Sitter boundary to $q \to -1.03$, whereas the $\text{CC}+\text{SN}^+$ and joint $\text{CC}+\text{SN}^++\text{BAO}$ trajectories remain above the de Sitter line in the quintessence-like regime at $q \to -0.92$ and $q \to -0.89$, respectively. When anchoring the expansion scale with the local SH0ES (2022) prior in FIG.~\ref{fig:powerlaw model dec with $H_0$ prior from SHOES-2020}, the extra power-law degree of freedom accommodates the higher local Hubble scale in $\text{CC}+H_0$ by driving stronger present-day acceleration at $q_0 \approx -0.68$ and pushing the far-future asymptote into the super-accelerating phantom regime at $q \to -1.21$. Conversely, incorporating intermediate-redshift supernova luminosity distances in $\text{CC}+\text{SN}^++H_0$ shifts the transition epoch earlier to $z_t \approx 0.50 $ with a milder present-day deceleration of $q_0 \approx -0.35$ and a future asymptote of $q \to -0.84$. Including BAO(DESI-DR2) sound-horizon constraints in the full joint fit $\text{CC}+\text{SN}^++\text{BAO}+H_0$ resolves this tension and pulls the trajectory back onto the $\Lambda\text{CDM}$ curve across the entire cosmic history, recovering $z_t \approx 0.65$, $q_0 \approx -0.54$, and a future asymptote just below the de Sitter limit at $q \to -1.01$.\\

\begin{figure*}[t!]
    \centering
    \begin{subfigure}[t]{0.48\textwidth}
        \centering
        \includegraphics[width=\linewidth]{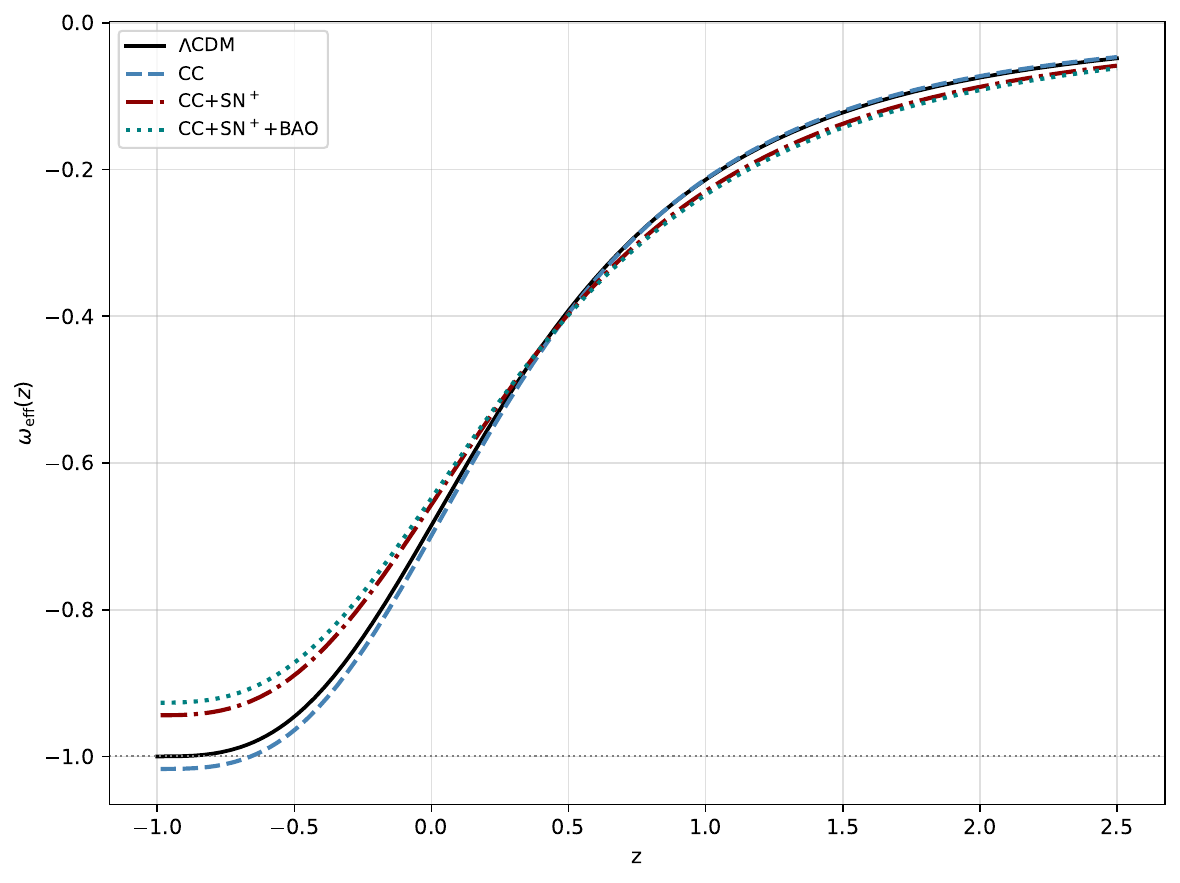}
        \caption{With uniform $H_0$ prior}
        \label{fig:power_law_eos}
    \end{subfigure}
    \hspace{0.1 cm}
    \hfill
    \begin{subfigure}[t]{0.48\textwidth}
        \centering
        \includegraphics[width=\linewidth]{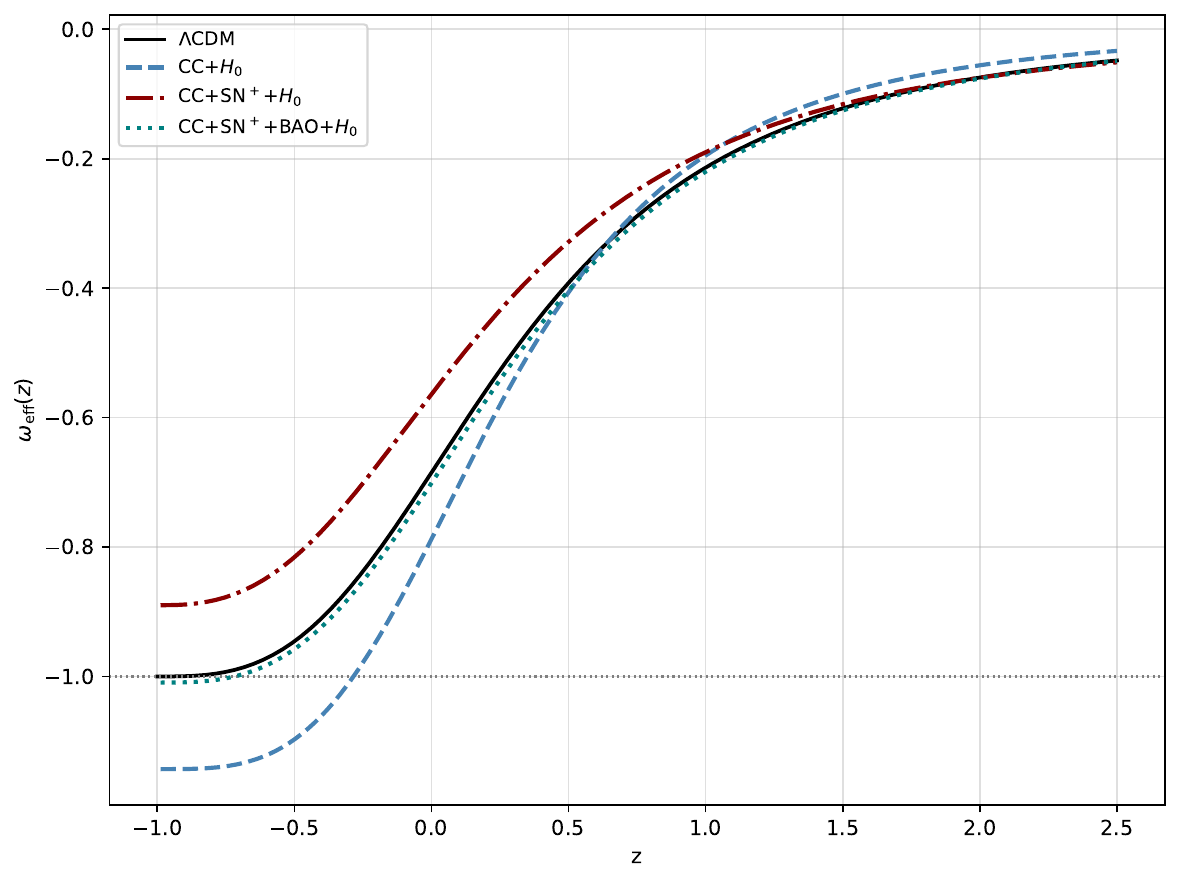}
        \caption{With Gaussian prior from SH0ES(2022) $H_0$ measurement}
        \label{fig:power_law_eosH0}
    \end{subfigure}
    \caption{Eos parameter vs redshift plots for \hyperref[modelI]{Model-I} with the constrained parameter values: FIG.~\ref{fig:power_law_eos} is with the uniform $H_0$ prior and FIG.~\ref{fig:power_law_eosH0} with a Gaussian $H_0$ prior from SH0ES-2022 measurement}
    \label{fig:combined_Eos(z)vsz_power-law}
\end{figure*}

 Under the uniform $H_0$ prior in FIG.~\ref{fig:power_law_eos}, $\omega_{\mathrm{eff}}(z)$ evolves monotonically within the quintessence-like regime $-1 < \omega_{\mathrm{eff}}(z) < -1/3$ throughout all past and present epochs, attaining present values of $\omega_{\mathrm{eff}}(0) \approx -0.70$ for CC and $\omega_{\mathrm{eff}}(0) \approx -0.65$ for $\text{CC}+\text{SN}^+$ and $\text{CC}+\text{SN}^++\text{BAO}$, before smoothly approaching the pressureless matter era $\omega_{\mathrm{eff}} \to 0$ at $z > 2.0$ and settling at $\omega_{\mathrm{eff}} \to -0.93$ for the joint data set as $z \to -1$. Under the local SH0ES(2022) anchor in FIG.~\ref{fig:power_law_eosH0}, the $\text{CC}+H_0$ reconstruction exhibits a steeper negative evolution, reaching $\omega_{\mathrm{eff}}(0) \approx -0.78$ in the present epoch, crossing the phantom divide $\omega_{\mathrm{eff}} = -1$ in the near future around $z \approx -0.32$, and asymptotically reaching $\omega_{\mathrm{eff}} \approx -1.14$ as $z \to -1$. In contrast, the $\text{CC}+\text{SN}^++H_0$ trajectory remains strictly quintessence-like with $\omega_{\mathrm{eff}}(0) \approx -0.56$ and a future asymptote of $\omega_{\mathrm{eff}} \to -0.89$. When (DESI-DR2)BAO constraints are incorporated in the full joint combination $\text{CC}+\text{SN}^++\text{BAO}+H_0$, both the excessive phantom excursion and quintessence shift are suppressed, locking the expansion history onto the cosmological constant trajectory with $\omega_{\mathrm{eff}}(0) \approx -0.69$ and crossing marginally into the phantom regime at $\omega_{\mathrm{eff}} \to -1.01$ as $z \to -1$.\\

\begin{figure*}[t!]
    \centering
    \begin{subfigure}[t]{0.48\textwidth}
        \centering
        \includegraphics[width=\linewidth]{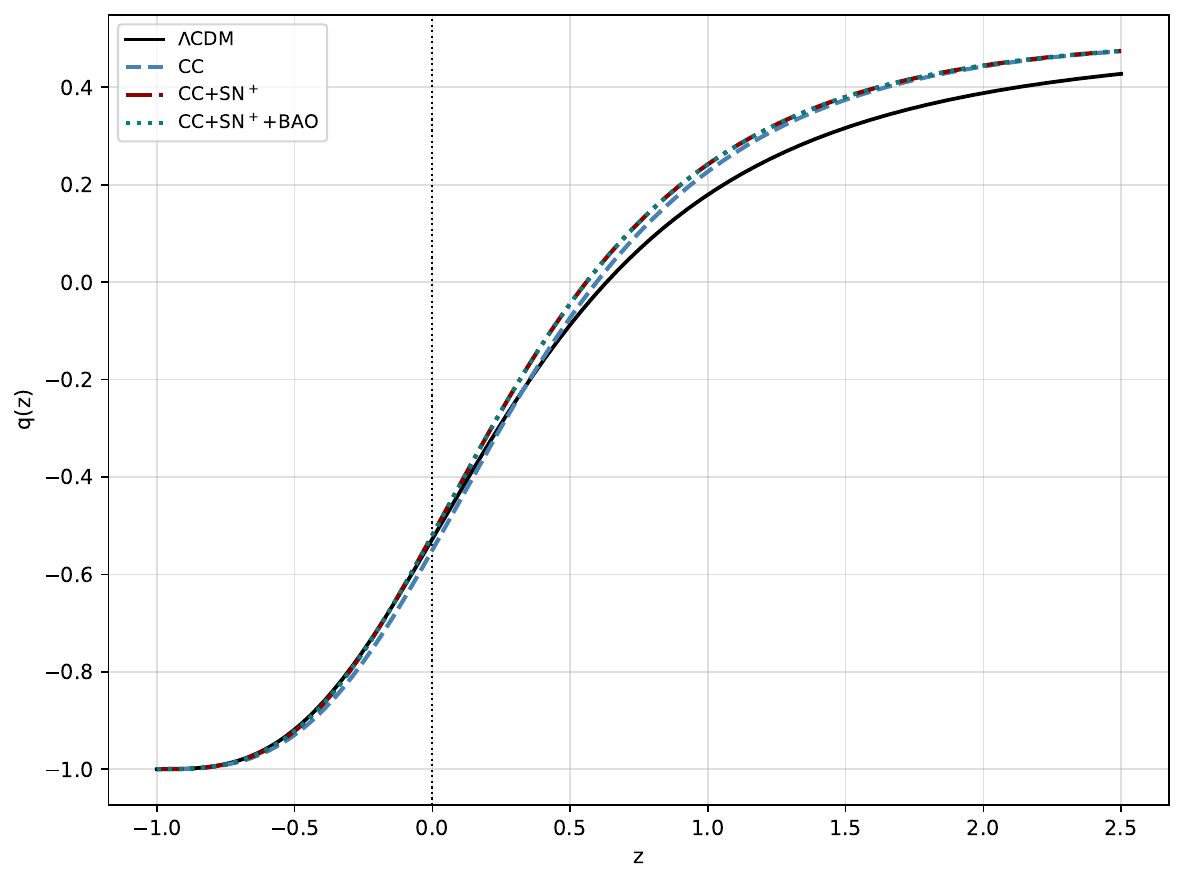}
        \caption{With uniform $H_0$ prior}
        \label{fig:exponential model dec without $H_0$ prior}
    \end{subfigure}
    \hspace{0.1 cm}
    \hfill
    \begin{subfigure}[t]{0.48\textwidth}
        \centering
        \includegraphics[width=\linewidth]{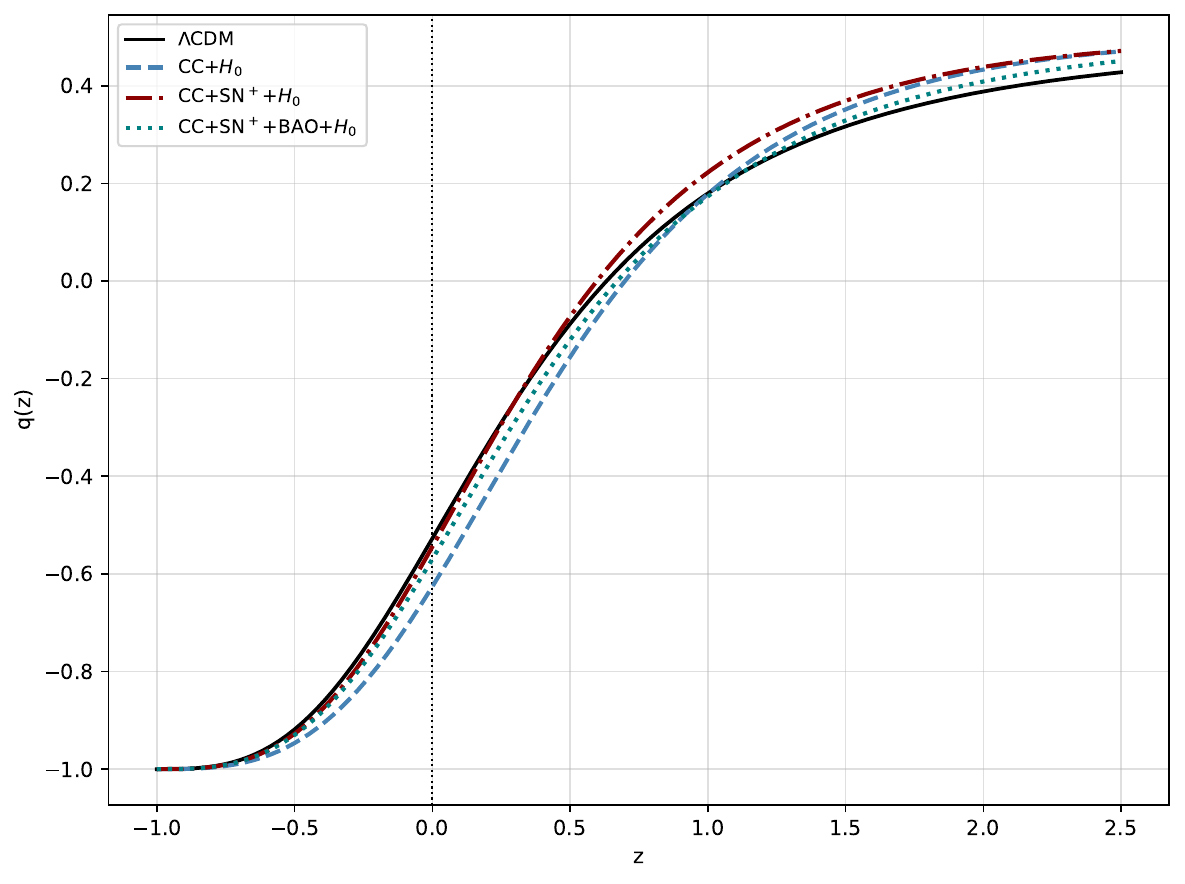}
        \caption{With Gaussian prior from SH0ES(2022) $H_0$ measurement}
        \label{fig:exponential model dec with $H_0$ prior from SHOES-2020}
    \end{subfigure}
    \caption{Deceleration parameter vs redshift plots for \hyperref[modelII]{Model-II}: FIG.~\ref{fig:exponential model dec without $H_0$ prior} is with the uniform $H_0$ prior and FIG.~\ref{fig:exponential model dec with $H_0$ prior from SHOES-2020} with a Gaussian $H_0$ prior from SH0ES-2022 measurement.}
\end{figure*}

For \hyperref[modelII]{Model-II} in the unanchored regime, $q(z)$ exhibits remarkable visual stability and tight clustering across all combinations of datasets, as shown in FIG.~\ref{fig:exponential model dec without $H_0$ prior}. The transition redshift occurs at $z_t \approx 0.56 - 0.60$, with the present-day deceleration parameter settling between $q_0 \approx -0.53$ and $-0.56$. At high redshifts $z > 1.5$, all three reconstructions converge smoothly toward the matter-dominated asymptote $q \to 0.5$, lying slightly above the $\Lambda\text{CDM}$ baseline, while in the far-future limit $z \to -1$ every trajectory approaches the exact de Sitter attractor $q \to -1$. When anchoring the expansion scale with the local SH0ES 2022 prior in FIG.~\ref{fig:exponential model dec with $H_0$ prior from SHOES-2020}, the $\text{CC}+H_0$ fit absorbs the higher local Hubble rate by shifting present-day acceleration to $q_0 \approx -0.63$ with a transition at $z_t \approx 0.67$, whereas incorporating intermediate-redshift supernova distances in $\text{CC}+\text{SN}^++H_0$ moderates the evolution to $q_0 \approx -0.54$ and $z_t \approx 0.56$. Adding DESI DR-2 BAO sound-horizon constraints in the full joint fit $\text{CC}+\text{SN}^++\text{BAO}+H_0$ mediates this trade-off smoothly at $q_0 \approx -0.58$ and $z_t \approx 0.62$ without inducing steep gradient anomalies, demonstrating greater dynamical stability than \hyperref[modelI]{Model-I}.\\

\begin{figure*}[t!]
    \centering
    \begin{subfigure}[t]{0.48\textwidth}
        \centering
        \includegraphics[width=\linewidth]{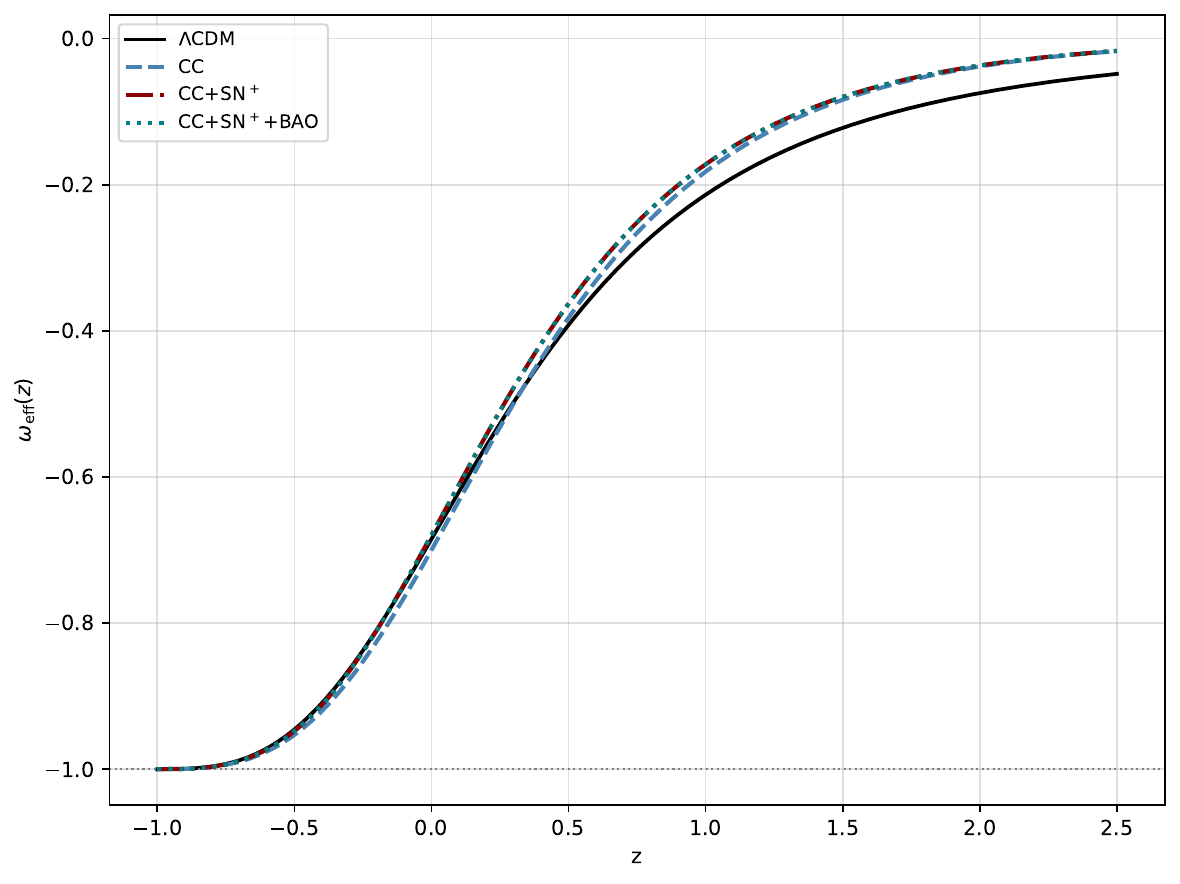}
        \caption{With uniform $H_0$ prior}
        \label{fig:exponential model eos without $H_0$ prior}
    \end{subfigure}
    \hspace{0.1 cm}
    \hfill
    \begin{subfigure}[t]{0.48\textwidth}
        \centering
        \includegraphics[width=\linewidth]{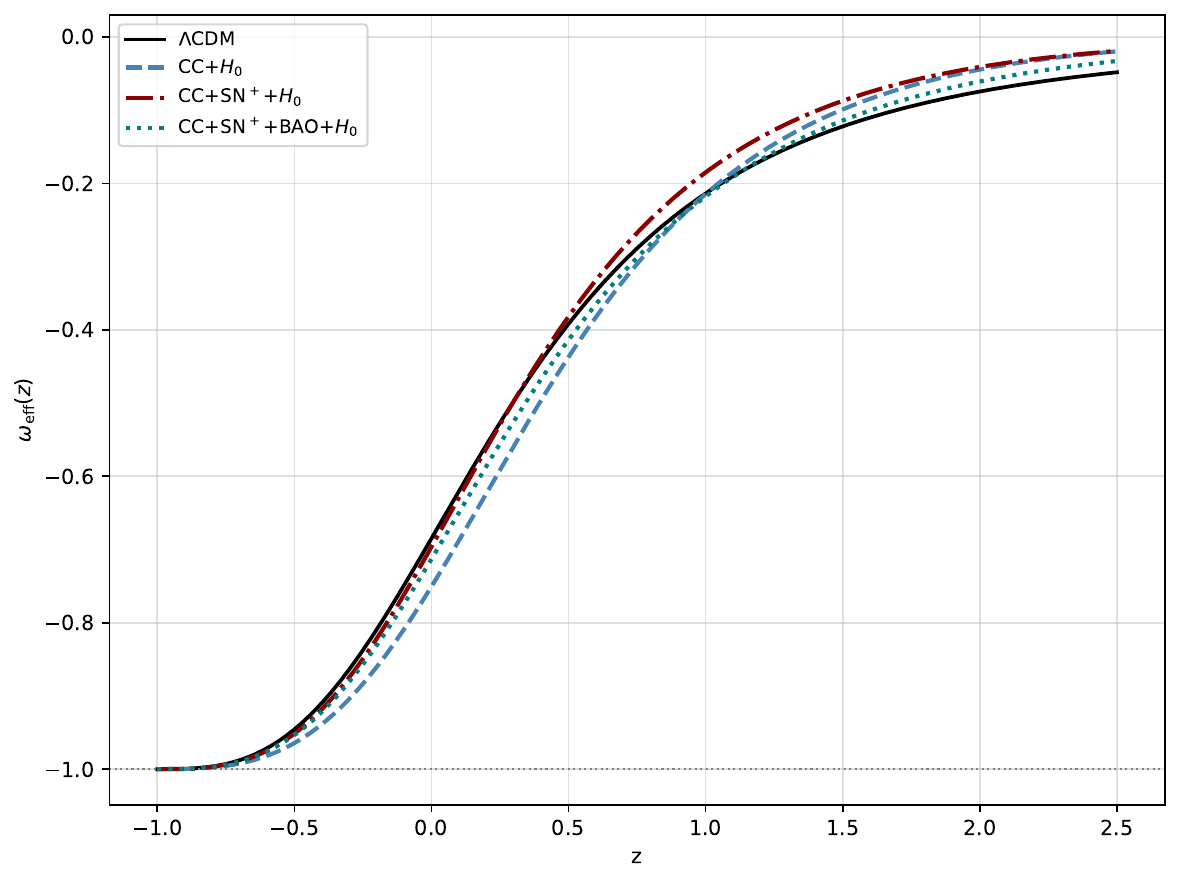}
        \caption{With Gaussian prior from SH0ES(2022) $H_0$ measurement}
        \label{fig:exponential model eos with $H_0$ prior from SHOES-2020}
    \end{subfigure}
    \caption {EoS parameter vs redshift plots for \hyperref[modelII]{Model-II}: Fig:\ref{fig:exponential model eos without $H_0$ prior} is with the uniform $H_0$ prior and FIG.~\ref{fig:exponential model eos with $H_0$ prior from SHOES-2020} with a Gaussian $H_0$ prior from SH0ES-2020 measurement.}
    \label{fig:combined_omega(z)vsz_exponential-model}
\end{figure*}

The corresponding evolution of the effective equation of state $\omega_{\mathrm{eff}}(z)$ for \hyperref[modelII]{Model-II} is presented in FIG.~\ref{fig:exponential model eos without $H_0$ prior} and FIG.~\ref{fig:exponential model eos with $H_0$ prior from SHOES-2020}. Under the uniform $H_0$ prior, the $\text{CC}$, $\text{CC}+\text{SN}^+$, and full joint $\text{CC}+\text{SN}^++\text{BAO}$ trajectories form an extremely tight bundle that remains strictly within the quintessence-like regime $-1 < \omega_{\mathrm{eff}}(z) < -1/3$ across all past and present epochs. At the present epoch, all three reconstructions closely match the $\Lambda\text{CDM}$ value at $\omega_{\mathrm{eff}}(0) \approx -0.68$ to $-0.70$, before converging monotonically toward the cosmological constant asymptote $\omega_{\mathrm{eff}} \to -1$ as $z \to -1$. At intermediate and high redshifts $z \gtrsim 0.5$, the trajectories lie slightly above the $\Lambda\text{CDM}$ baseline and approach the pressureless matter-dominated limit $\omega_{\mathrm{eff}} \to 0$ more rapidly, reaching $\omega_{\mathrm{eff}} \approx -0.02$ by $z = 2.5$. When anchoring the expansion scale with the local SH0ES prior in FIG.~\ref{fig:exponential model eos with $H_0$ prior from SHOES-2020}, the model preserves high dynamical stability without crossing the phantom divide $\omega_{\mathrm{eff}} = -1$. Specifically, the $\text{CC}+H_0$ reconstruction exhibits a mildly more negative present-day equation of state at $\omega_{\mathrm{eff}}(0) \approx -0.75$ to accommodate the elevated local expansion rate, whereas $\text{CC}+\text{SN}^++H_0$ shifts slightly upward at intermediate redshifts, and the full joint combination $\text{CC}+\text{SN}^++\text{BAO}+H_0$ closely tracks the $\Lambda\text{CDM}$ curve with $\omega_{\mathrm{eff}}(0) \approx -0.71$ and a smooth asymptotic approach to $\omega_{\mathrm{eff}} \to -1$ as $z \to -1$. Thus, the tension between the local $H_0$ anchor and intermediate-redshift BAO geometry in \hyperref[modelII]{Model-II} does not induce late-time phantom instabilities, manifesting instead as a mild Bayesian complexity penalty of $\Delta\text{BIC} = 1.17$.\\

Across all unanchored fits, the model $\omega_{\mathrm{eff}}(z)$ curves form an extremely tight bundle FIG. \ref{fig:exponential model eos without $H_0$ prior}. The model remains strictly within the quintessence domain at late times, evolving smoothly toward $\omega_{\mathrm{eff}} \to -1$ as $z \to -1$ and vanishing $\omega_{\mathrm{eff}} \to 0$ during the matter-dominated epoch $z > 1.5$. Under local $H_0$ anchoring, the model maintains high stability FIG.~\ref{fig:exponential model eos with $H_0$ prior from SHOES-2020}. Unlike model--I, it does not suffer from phantom-like dips; instead,the exponential coupling provides mildly stronger late-time negative pressure in $\text{CC}+H_0$ at $\omega_{\mathrm{eff}}(0) \approx -0.75$ to elevate $H_0$, while adding $\text{SN}^+$ and (DESI-DR2) BAO constraints pulls the joint trajectory back onto the $\Lambda\text{CDM}$ curve with $\omega_{\mathrm{eff}}(0) \approx -0.71$ and $\omega_{\mathrm{eff}} \to -1$ as $z \to -1$. Consequently, the tension between the local $H_0$ anchor and intermediate-redshift BAO geometry manifests not as phantom-divide crossings, but rather as a mild Bayesian complexity penalty of $\Delta\text{BIC} = 1.17$ alongside $\Delta\text{AIC} = -4.29$ and $\Delta\text{DIC} = -1.47$.

\section{Conclusion}\label{VI}

We have investigated the late-time cosmological dynamics and observational viability of two non-minimal curvature--matter coupling cosmological models in ${f(R, \mathcal{L}_m})$ gravity with two functional forms, such as power-law form and exponential form. To constrain model parameters ($b_1$ and $b_2$) and cosmological parameters ($H_0$, $\Omega_m$), we performed full $MCMC$ sampling using the Cobaya framework across two distinct operational regimes: an unanchored scenario with a flat uniform $H_0$ prior, and a locally anchored scenario incorporating the $H_0$ prior. Both regimes were tested against three cumulative observational dataset combinations such as $\text{CC}$, $\text{CC} + \text{SN}^+$, and $\text{CC} + \text{SN}^+ + \text{BAO(DESI-DR2)}$. To rigorously quantify model performance and assess the trade-off between statistical fit quality and model complexity, we conducted a comprehensive statistical analysis evaluating $\text{AIC}$, $\text{BIC}$, $\text{DIC}$ and their respective differences ($\Delta\text{AIC}$, $\Delta\text{BIC}$, $\Delta\text{DIC}$) relative to the standard flat baseline $\Lambda\text{CDM}$. In the unanchored joint analysis ($\text{CC}+\text{SN}^{+}+\text{BAO}$), the baseline flat $\Lambda\text{CDM}$ model yields $H_{0} = 68.24 \pm 0.39\,\mathrm{km\,s^{-1}\,Mpc^{-1}}$, which stands in $4.3\sigma$ tension with the local SH0ES 2022 measurement. Under the same unanchored dataset combination, \hyperref[modelI]{Model-I} constrains $H_{0} = 67.00 \pm 1.04\,\mathrm{km\,s^{-1}\,Mpc^{-1}}$ with $b_{1} = 0.073 \pm 0.037$, shifting the central value downward relative to $\Lambda\text{CDM}$ and leaving the Hubble tension largely unchanged at $4.1\sigma$. In contrast, \hyperref[modelII]{Model-II} yields $H_{0} = 69.46 \pm 0.70\,\mathrm{km\,s^{-1}\,Mpc^{-1}}$ with $b_{2} = -0.004 \pm 0.122$, shifting the inferred Hubble constant upward by $+1.22\,\mathrm{km\,s^{-1}\,Mpc^{-1}}$ compared to $\Lambda\text{CDM}$ and modestly alleviating the Hubble tension to $2.9\sigma$ without imposing any external distance-ladder prior. When the local SH0ES 2022 prior is introduced, both models accommodate $H_{0} \approx 73.0\,\mathrm{km\,s^{-1}\,Mpc^{-1}}$ in the $\text{CC}+H_{0}$; however, incorporating intermediate-redshift $\text{SN}^{+}$ distances and (DESI-DR2)BAO sound-horizon constraints pulls the joint expansion rates down to $H_{0} = 69.44 \pm 0.80\,\mathrm{km\,s^{-1}\,Mpc^{-1}}$ ($2.7\sigma$ discrepancy) for \hyperref[modelI]{Model-I} and $H_{0} = 70.60 \pm 0.86\,\mathrm{km\,s^{-1}\,Mpc^{-1}}$ ($1.8\sigma$ discrepancy) for \hyperref[modelII]{Model-II}, compared to $H_{0} = 69.25 \pm 0.37\,\mathrm{km\,s^{-1}\,Mpc^{-1}}$ ($3.4\sigma$) for $\Lambda\text{CDM}$. Thus, while neither model fully eliminates the geometric conflict between local distance-ladder calibration and BAO sound-horizon scales, \hyperref[modelII]{Model-II} consistently provides greater relief of the $H_{0}$ tension than both $\Lambda\text{CDM}$ and \hyperref[modelI]{Model-I}.\\ 
The kinematic deceleration parameter $q(z)$ confirms that both frameworks execute a smooth transition from early matter-dominated deceleration ($q > 0$) to late-time cosmic acceleration ($q < 0$), recovering the standard Einstein--de Sitter limit $q \to 0.5$ at high redshifts $z > 2.0$. In the unanchored regime, the transition occurs at $z_{t} \approx 0.62\text{--}0.67$ with $q_{0} \approx -0.48$ to $-0.55$ for \hyperref[modelI]{Model-I}, and at $z_{t} \approx 0.56\text{--}0.60$ with $q_{0} \approx -0.53$ to $-0.56$ for \hyperref[modelII]{Model-II}. Analysis of the effective equation of state $\omega_{\mathrm{eff}}(z)$ highlights a clear dynamical distinction between the two parameterizations. Under uniform $H_{0}$ priors, both models evolve within the quintessence-like domain $-1 < \omega_{\mathrm{eff}}(z) < -1/3$ across all past and present epochs. Under local SH0ES anchoring, however, \hyperref[modelI]{Model-I} exhibits pronounced dataset sensitivity: its $\text{CC}+H_{0}$ trajectory crosses the phantom divide at $z \approx -0.32$ toward $\omega_{\mathrm{eff}} \to -1.14$ ($q \to -1.21$) as $z \to -1$, whereas adding DESI DR2 BAO forces $b_{1} \to -0.009 \pm 0.030$ and pulls the trajectory back onto $\Lambda\text{CDM}$. Conversely, \hyperref[modelII]{Model-II} maintains exceptional dynamical robustness across both prior regimes, keeping $\omega_{\mathrm{eff}}(z)$ strictly above the phantom divide and converging uniformly to the exact de Sitter attractor ($q \to -1$, $\omega_{\mathrm{eff}} \to -1$) in the asymptotic future.\\
Information-theoretic model selection underscores this structural contrast. In the unanchored joint fit ($\text{CC}+\text{SN}^{+}+\text{BAO}$), the improved likelihood allows both models to overcome Occam's complexity penalty, yielding strong statistical support for \hyperref[modelI]{Model-I} ($\Delta\text{AIC} = -12.07$, $\Delta\text{BIC} = -6.60$, $\Delta\text{DIC} = -12.17$) and positive-to-strong support for \hyperref[modelII]{Model-II} ($\Delta\text{AIC} = -8.37$, $\Delta\text{BIC} = -2.91$, $\Delta\text{DIC} = -9.55$). Under the SH0ES-anchored joint dataset ($\text{CC}+\text{SN}^{+}+\text{BAO}+H_{0}$), \hyperref[modelI]{Model-I} degenerates back to $\Lambda\text{CDM}$ and suffers a strong Bayesian penalty ($\Delta\text{AIC} = +1.95$, $\Delta\text{BIC} = +7.42$, $\Delta\text{DIC} = +1.67$). In contrast, \hyperref[modelII]{Model-II} partially absorbs the $H_{0}$ tension to achieve positive frequentist support ($\Delta\text{AIC} = -4.29$), while Occam's $\ln N$ penalty places the Bayesian metric in the inconclusive regime ($\Delta\text{BIC} = +1.17$) and the bimodal posterior variance of $b_{2}$ moderates $\Delta\text{DIC}$ to $-1.47$. Overall, \hyperref[modelII]{Model-II} outperforms \hyperref[modelI]{Model-I} in both dynamical stability and SH0ES-anchored statistical competitiveness, offering a physically consistent late-time mechanism that modestly eases the $H_{0}$ tension relative to $\Lambda\text{CDM}$ while remaining statistically competitive with the standard cosmological baseline.

\section{Appendix}
\subsection{Appendix-1}\label{Appendix-1}
In this appendix, we provide the derivation of the background Friedmann equations for the power-law  $f(R,\mathcal{L}_m)$ model considered in this work. Starting from the field equations of $f(R,\mathcal{L}_m)$ gravity given in Eq.~\eqref{fieldequation1111} and Eq.~\eqref{fieldequation2} and subsequently specialize them to the power-law forms of
$f(R,\mathcal{L}_m)$.

\begin{equation}
f(R,\mathcal{L}_m)= \frac{R}{2}+\mathcal{L}_m+\alpha \mathcal{L}_m^{b_1}, \quad f_R=\frac{1}{2}, \quad
f_{\mathcal{L}_m}=1+\alpha b_1\mathcal{L}_m^{b_1-1}.
\end{equation}

For the pressureless matter component considered here, $p_m=0$, and the matter density evolves as $\rho_m(z)=\rho_{m0}(1+z)^3.$ This relation is adopted to construct a background expansion history that can be confronted with cosmological observations. In the $f(R,\mathcal{L}_m)$ framework, the standard dust-scaling relation can be recovered under specific conditions, when the modified continuity equation is compatible. For the power-law model considered here, this compatibility is obtained for \(b_1=0\), \(b_1=1\), or \(\alpha=0\), under the regularity assumptions of the model. For general parameter values, the standard scaling is imposed as a phenomenological prescription, and the resulting observational constraints are interpreted within the corresponding background parametrization.  The present-day matter density is related to the matter density parameter
through $\rho_{m0}=3H_0^2\Omega_{m0}$. Therefore, $\rho_m(z)=3H_0^2\Omega_{m0}(1+z)^3$. So for the power-law model, Eq.~\eqref {fieldequation1111} is reduced as:
\begin{equation}\label{Hrho}
3H^2=\rho_m+\alpha(2b_1-1)\rho_m^{b_1}.
\end{equation}

Substituting $\rho_m(z)$ in Eq.~\eqref{Hrho} and \eqref{fieldequation2}, the above equation becomes
\begin{align}
3H^2(z)&=3H_0^2\Omega_{m0}(1+z)^3+\alpha(2b_1-1) \left[3H_0^2\Omega_{m0}(1+z)^3\right]^{b_1},\label{p1}\\
-2\dot{H(z)}-3H^2(z)&= -\alpha(b_1-1)\left[3H_0^2\Omega_{m0}(1+z)^3\right]^{b_1}
\label{p2}
\end{align}

Evaluating Eq.~\eqref{p1} at present time, we can express $\alpha$ in terms of $H_0$,$\Omega_{m0}$ and $b_1$

\begin{equation}\label{alpha}
    \alpha=\frac{1-\Omega_{m0}}{\Omega_{m0}(2b_1-1)\left(3H_0^2\Omega_{m0}\right)^{b_1}}
\end{equation}
Substituting Eq.~\eqref{alpha} back into Eq.~\eqref{p1}, we express it as a function of redshift 'z'  where $E(z)=\frac{H(z)}{H_0}$ as:
\begin{equation*}
    E(z)=\left[\Omega_{m0}(1+z)^3+(1-\Omega_{m0})(1+z)^{3b_1}\right]^{\frac{1}{2}}
\end{equation*}
Thus, after imposing the present normalization, the background expansion history of Model I is determined by the parameters $H_0$, $\Omega_{m0}$, and $b_1$.

\subsection{Appendix-II}\label{Appendix-II}
 As in \hyperref[modelI]{Model-I}, we consider pressureless matter with $\mathcal{L}_m=\rho_m$ and initially adopt the standard dust-scaling relation $\rho_m(z)=\rho_{m0}(1+z)^3$, where $\rho_{m0}=3H_0^2\Omega_{m0}$. This assumption is used to derive a phenomenological background expansion for comparison with observations. It is not generally guaranteed by the modified continuity equation, because the exponential matter coupling makes $f_{\mathcal{L}_m}$ density-dependent. We therefore distinguish the background expressions obtained under this assumption from a fully self-consistent solution of the modified field equations. For the exponential non-minimal coupling model, we consider:
\begin{equation}
f(R,\mathcal{L}_m)= \frac{R}{2}+\mathcal{L}_m-\beta \mathcal{L}_{m0}e^{-b_2\sqrt{\frac{\mathcal{L}_m}{\mathcal{L}_{m0}}}}, \quad f_R=\frac{1}{2}, \quad
f_{\mathcal{L}_m}=1+\frac{\beta b_2}{2}\sqrt{\frac{\rho}{\rho_0}}e^{-b_2(1+z)^\frac{3}{2}}.
\end{equation}

The primary physical motivation for adopting this exponential non-minimal matter-geometry coupling is to explore whether non-linear modifications in the matter sector can naturally drive late-time cosmic acceleration without invoking an ad hoc dark energy scalar field or cosmological constant. In this framework, the exponential modification introduces a density-dependent damping factor that becomes dynamically dominant at low redshifts, offering a self-consistent mechanism for late-time cosmic acceleration.  Evaluating the modified field equations under this background scaling for exponential power model, Eq.~\eqref {fieldequation1111} is reduced as:

\begin{equation}\label{Hexporho}
3H^2=\rho_m+\beta \rho_0\left[(1+b_2(1+z)^\frac{3}{2})e^{-b_2(1+z)^\frac{3}{2}}\right].
\end{equation}

Substituting $\rho_m(z)$ in Eq.~\eqref{Hexporho} and \eqref{fieldequation2}, the above equation becomes
\begin{align}
3H^2(z)= 3H_0^2[\Omega_{m0}(1+z)^3+\beta \Omega_{m0}[(1+b_2(1+z)^{\frac{3}{2}}]e^{-b_2(1+z)^{\frac{3}{2}}}], \label{e1}\\
-3H^2(z)-2\dot{H}(z)=-3H_0^2\Omega_{m0}[(1+\frac{b_2(1+z)^{\frac{3}{2}}}{2})\beta e^{-b_2(1+z)^{\frac{3}{2}}}] \label{e2}.
\end{align}

Evaluating Eq.~\eqref{e1} at present time, we can express the $\beta$ in terms of $H_0$,$\Omega_{m0}$ and $b_2$

\begin{equation}\label{beta}
    \beta=\frac{1-\Omega_{m0}}{\Omega_{m0}[(1+b_2)e^{-b_2}]}
\end{equation}
Substituting Eq.~\eqref{beta} back into Eq.~\eqref{e1}, we express it as a function of redshift 'z', where $E(z)=\frac{H(z)}{H_0}$ as:
\begin{equation*}
   E(z)=\left[\Omega_{m0}(1+z)^3+(1-\Omega_{m0})[\frac{(1+b_2(1+z)^{\frac{3}{2}})e^{-b_2(1+z)^{\frac{3}{2}}}}{(1+b_2)e^{-b_2}}]\right]^{\frac{1}{2}}.
\end{equation*}
Thus, after imposing the present normalization, the background expansion history of Model II is determined by the parameters $H_0$, $\Omega_{m0}$, and $b_2$.

\subsection{Appendix-III}
The posterior distributions of $H_0$ and $\Omega_{m0}$ for the $\Lambda$CDM model are shown for the considered datasets, with and without the $H_0$ prior from SH0ES(2022). The corresponding parameter constraints and statistical measures, including $\chi^2_{\min}$, AIC, BIC, and DIC, are summarized in Table~\ref{tab:aic_bic} for the different combinations of CC, $\mathrm{SN}^+$, and BAO datasets.

\begin{figure}[htbp]
    \centering
    \begin{subfigure}[b]{0.43\textwidth}
        \centering
        \includegraphics[width=1.2\textwidth]{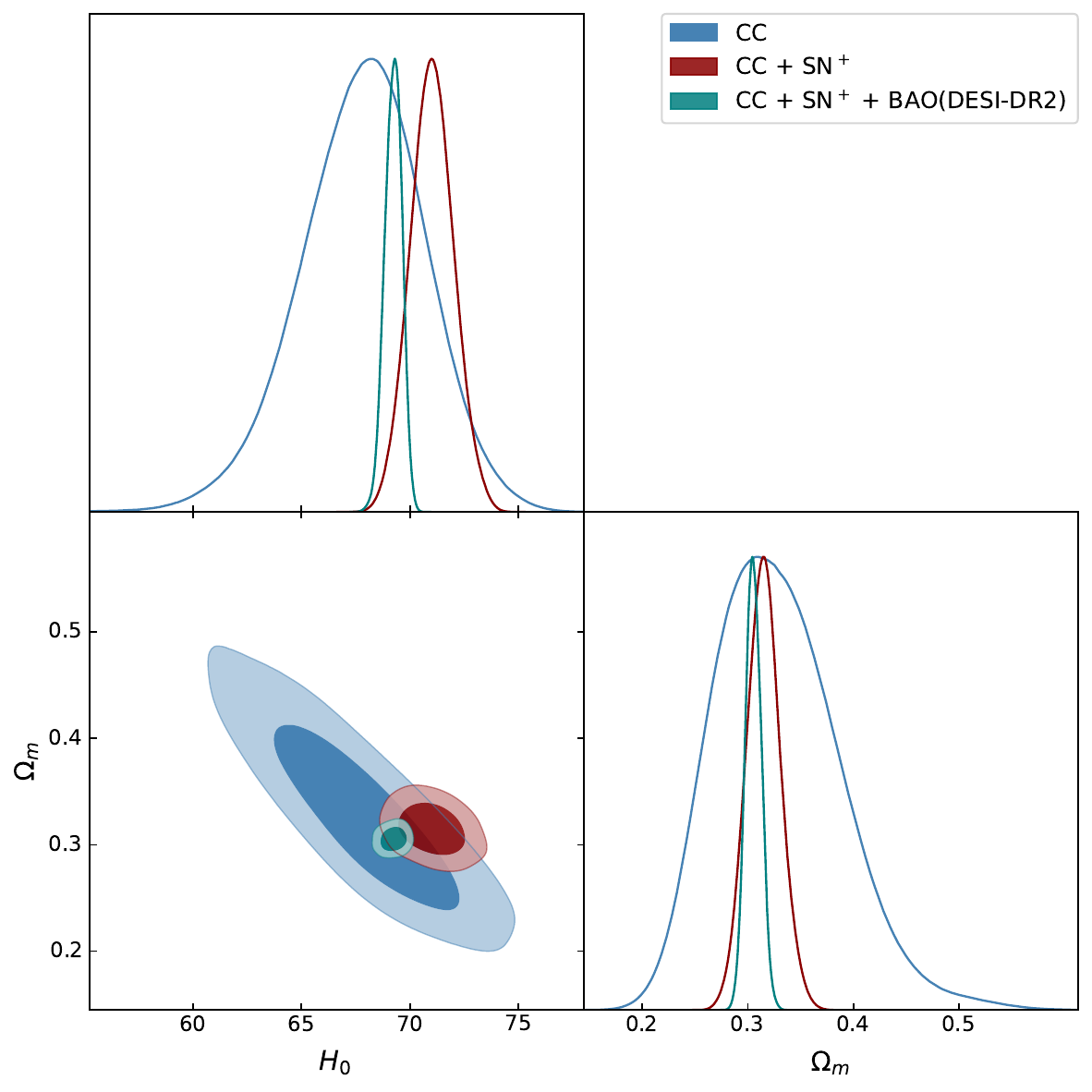}
        \caption{}
        \label{fig:exponential model without $H_0$ prior}
    \end{subfigure}
    \hspace{0.1 cm}
    \hfill
    \begin{subfigure}[b]{0.43\textwidth}
        \centering
        \includegraphics[width=1.2\textwidth]{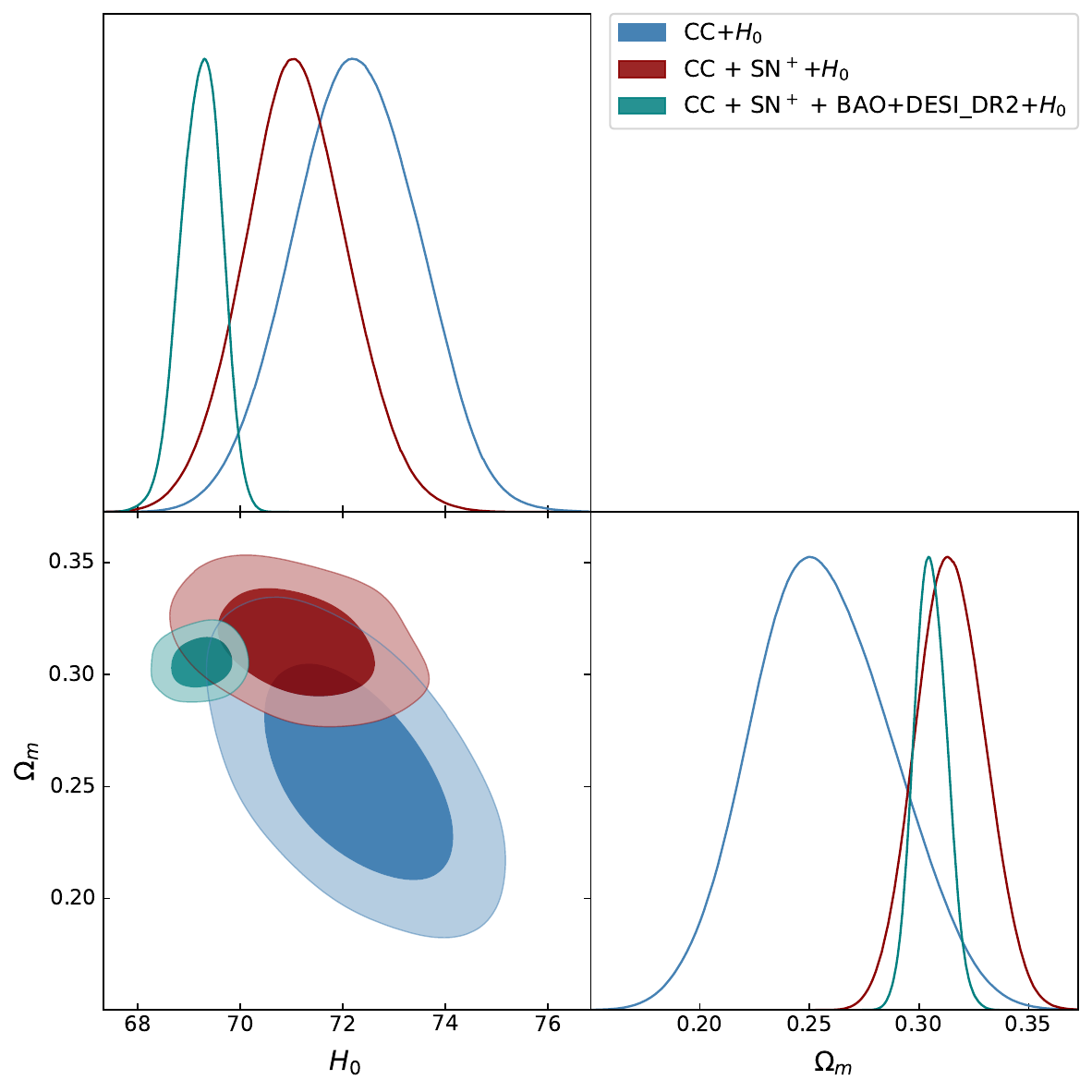}
        \caption{}
        \label{fig:Lambda CDM with H_0 prior  }
    \end{subfigure}
    \caption{Constrained $\Lambda$CDM cosmological parameters with and without $H_0$ priors}
    \label{}
\end{figure}

{\color{blue}
\begin{table}[htbp]
\footnotesize
\centering
\caption{}
\label{tab:aic_bic}
\begin{tabular}{lcccccc}
\hline
Dataset & $H_0$ & $\Omega_{m0}$ & $\chi^2_{\min}$ & AIC & BIC & DIC    \\
\hline
\multicolumn{7}{c}{With uniform $H_0$ prior} \\
\hline
CC & 67.83702 $\pm$ 2.85622   & 0.32796  $\pm$ 0.05943 &  14.5241 & 18.9527 & 21.3921 & 18.4071  \\
CC + SN$^+$  & 71.01737  $\pm$ 0.99415 & 0.31473 $\pm$ 0.01611 & 1424.5579 & 1428.5579 & 1439.4720 & 1428.5788 \\
CC + SN$^+$ + BAO & 68.23502 $\pm$ 0.39173 & 0.30530 $\pm$ 0.00738 & 1440.9037 & 1444.9037 & 1455.8270 & 1445.1043 \\
\hline
\multicolumn{7}{c}{With $H_0$ (SHOES 2022)} \\
\hline
CC & 72.25730  $\pm$ 1.19582 & 0.27439  $\pm$ 0.03113 & 17.1653
 & 21.5791 & 24.0968 & 21.336 \\
CC + SN$^+$  &  71.09947  $\pm$ 1.00666 & 0.31437 $\pm$ 0.01542 & 1424.5575 & 1428.5575 & 1439.4727 & 1428.3892 \\
CC + SN$^+$  + BAO & 69.24761  $\pm$ 0.37167 & 0.30527  $\pm$ 0.00760 & 1440.9057 & 1444.9057 & 1455.8301 & 1444.9747  \\
\hline
\end{tabular}
\end{table}}
\section*{Acknowledgement} BM acknowledges the support of Council of Scientific and Industrial Research (CSIR) for the project grant (No. 03/1493/23/EMR II). YKD and BM acknowledge IUCAA, Pune (India) for providing support in the form of an academic visit during which part of this work is accomplished. 
\section*{References}
\bibliographystyle{JCAP.bst}
\bibliography{Refs}

@article{Harko_2010b,
  title={{$f(R, L_{m})$ gravity}},
  author={Harko, Tiberiu and Lobo, Francisco SN},
  journal={Eur. Phys. J. C},
  volume={70},
  pages={373},
  year={2010},
  doi = {10.1140/epjc/s10052-010-1467-3},
  url = {https://doi.org/10.1140/epjc/s10052-010-1467-3}
}

@book{LandauLifshitz1975,
  author    = {Landau, L. D. and Lifshitz, E. M.},
  title     = {The Classical Theory of Fields},
  edition   = {4},
  series    = {Course of Theoretical Physics},
  volume    = {2},
  year      = {1975},
  publisher = {Butterworth-Heinemann},
  address   = {Oxford}
}

@article{Harko_2010,
  title = {{Galactic rotation curves in modified gravity with nonminimal coupling between matter and geometry}},
  author = {Harko, T.},
  journal = {Phys. Rev. D},
  volume = {81},
  pages = {084050},
  year = {2010},
  doi = {10.1103/PhysRevD.81.084050},
  url = {https://link.aps.org/doi/10.1103/PhysRevD.81.084050}
}

@article{Bhagat_2026_adf2,
author = {Bhagat, Rahul and Fomin, I. V. and Mishra, B.},
title = {{Effects of Matter Lagrangian in f(Q,T) Gravity: The Accelerating Cosmological Model}},
journal = {Annalen der Physik},
volume = {538},
number = {3},
pages = {e00001},
doi = {https://doi.org/10.1002/andp.202600001},
year = {2026},
}

@article{devi2025late,
title = {Late time behavior in ${f(R,{L}_{m})}$ gravity through Gaussian reconstruction and dynamical stability},
journal = {Physics of the Dark Universe},
volume = {50},
pages = {102172},
year = {2025},
issn = {2212-6864},
doi = {https://doi.org/10.1016/j.dark.2025.102172},
url = {https://www.sciencedirect.com/science/article/pii/S2212686425003656},
author = {Y. Kalpana Devi and S. A. Narawade and B. Mishra},}

@article{Torrado_2021,
   title={{Cobaya: code for Bayesian analysis of hierarchical physical models}},
   volume={2021},
   ISSN={1475-7516},
   url={http://dx.doi.org/10.1088/1475-7516/2021/05/057},
   DOI={10.1088/1475-7516/2021/05/057},
   number={05},
   journal={Journal of Cosmology and Astroparticle Physics},
   publisher={IOP Publishing},
   author={Torrado, Jesús and Lewis, Antony},
   year={2021},
   month={May}, pages={057} }

@article{Jimenez_2002,
doi = {10.1086/340549},
url = {https://doi.org/10.1086/340549},
year = {2002},
month = {jul},
publisher = {},
volume = {573},
number = {1},
pages = {37},
author = {Jimenez, Raul and Loeb, Abraham},
title = {Constraining Cosmological Parameters Based on Relative Galaxy Ages},
journal = {The Astrophysical Journal}
}

@article{Moresco_2016,
  title={A 6\% measurement of the Hubble parameter at $z\sim0.45$: direct evidence of the epoch of cosmic re-acceleration},
  volume={2016},
  ISSN={1475-7516},
  url={http://dx.doi.org/10.1088/1475-7516/2016/05/014},
  DOI={10.1088/1475-7516/2016/05/014},
  number={05},
  journal={Journal of Cosmology and Astroparticle Physics},
  publisher={IOP Publishing},
  author={Moresco, Michele and Pozzetti, Lucia and Cimatti, Andrea and Jimenez, Raul and Maraston, Claudia and Verde, Licia and Thomas, Daniel and Citro, Annalisa and Tojeiro, Rita and Wilkinson, David},
  year={2016},
  month={May},
  pages={014--014}
}

@article{Scolnic_2022,
   title={The Pantheon+ Analysis: The Full Data Set and Light-curve Release},
   volume={938},
   ISSN={1538-4357},
   url={http://dx.doi.org/10.3847/1538-4357/ac8b7a},
   DOI={10.3847/1538-4357/ac8b7a},
   number={2},
   journal={The Astrophysical Journal},
   publisher={American Astronomical Society},
   author={Scolnic, Dan and Brout, Dillon and Carr, Anthony and Riess and others},
   year={2022},
   month=Oct, pages={113} 
   }

@article{Scolnic_2018,
   title={{The Complete Light-curve Sample of Spectroscopically Confirmed SNe Ia from Pan-STARRS1 and Cosmological Constraints from the Combined Pantheon Sample}},
   volume={859},
   ISSN={1538-4357},
   url={http://dx.doi.org/10.3847/1538-4357/aab9bb},
   DOI={10.3847/1538-4357/aab9bb},
   number={2},
   journal={The Astrophysical Journal},
   publisher={American Astronomical Society},
   author={Scolnic, D. M. and Jones, D. O. and Rest, A. and Pan, Y. C. and Chornock, R. and Foley, R. J. and Huber, M. E. and Kessler, R. and Narayan, G. and Riess, A. G. and Rodney, S. and Berger, E. and Brout, D. J. and Challis, P. J. and Drout, M. and Finkbeiner, D. and Lunnan, R. and Kirshner, R. P. and Sanders, N. E. and Schlafly, E. and Smartt, S. and Stubbs, C. W. and Tonry, J. and Wood-Vasey, W. M. and Foley, M. and Hand, J. and Johnson, E. and Burgett, W. S. and Chambers, K. C. and Draper, P. W. and Hodapp, K. W. and Kaiser, N. and Kudritzki, R. P. and Magnier, E. A. and Metcalfe, N. and Bresolin, F. and Gall, E. and Kotak, R. and McCrum, M. and Smith, K. W.},
   year={2018},
   month=May, 
   pages={101} 
   }

@article{Brout_2022,
   title={The Pantheon+ Analysis: Cosmological Constraints},
   volume={938},
   ISSN={1538-4357},
   url={http://dx.doi.org/10.3847/1538-4357/ac8e04},
   DOI={10.3847/1538-4357/ac8e04},
   number={2},
   journal={The Astrophysical Journal},
   publisher={American Astronomical Society},
   author={Brout, Dillon and Scolnic, Dan and Popovic, Brodie and Riess, Adam G. and Carr, Anthony and Zuntz, Joe and Kessler, Rick and Davis, Tamara M. and Hinton, Samuel and Jones, David and Kenworthy, W. D’Arcy and Peterson, Erik R. and Said, Khaled and Taylor, Georgie and Ali, Noor and Armstrong, Patrick and Charvu, Pranav and Dwomoh, Arianna and Meldorf, Cole and Palmese, Antonella and Qu, Helen and Rose, Benjamin M. and Sanchez, Bruno and Stubbs, Christopher W. and Vincenzi, Maria and Wood, Charlotte M. and Brown, Peter J. and Chen, Rebecca and Chambers, Ken and Coulter, David A. and Dai, Mi and Dimitriadis, Georgios and Filippenko, Alexei V. and Foley, Ryan J. and Jha, Saurabh W. and Kelsey, Lisa and Kirshner, Robert P. and Möller, Anais and Muir, Jessie and Nadathur, Seshadri and Pan, Yen-Chen and Rest, Armin and Rojas-Bravo, Cesar and Sako, Masao and Siebert, Matthew R. and Smith, Mat and Stahl, Benjamin E. and Wiseman, Phil},
   year={2022},
   month=Oct, pages={110} }

@article{Riess_2019,
   title={The expansion of the Universe is faster than expected},
   volume={2},
   ISSN={2522-5820},
   url={http://dx.doi.org/10.1038/s42254-019-0137-0},
   DOI={10.1038/s42254-019-0137-0},
   number={1},
   journal={Nature Reviews Physics},
   publisher={Springer Science and Business Media LLC},
   author={Riess, Adam G.},
   year={2019},
   month=Dec, pages={10-12} 
   }

@article{Seikel_2012,
  doi = {10.1088/1475-7516/2012/06/036},
  url = {https://dx.doi.org/10.1088/1475-7516/2012/06/036},
  year = {2012},
  volume = {2012},
  number = {06},
  pages = {036},
  author = {Marina Seikel and Chris Clarkson and Mathew Smith},
  title = {{Reconstruction of dark energy and expansion dynamics using Gaussian processes}},
  journal = {J. Cosmol. Astropart. Phys.}
}

@article{Riess_2018,
  doi = {10.3847/1538-4357/aaa5a9},
  url = {https://dx.doi.org/10.3847/1538-4357/aaa5a9},
  year = {2018},
  volume = {853},
  number = {2},
  pages = {126},
  author = {Adam G. Riess and Steven A. Rodney and Daniel M. Scolnic and Daniel L. Shafer},
  title = {{Type Ia Supernova Distances at Redshift $>$1.5 from the Hubble Space Telescope Multi-cycle Treasury Programs: The Early Expansion Rate}},
  journal = {Astrophys. J.}
}

@article{Zhang_2016,
  doi = {10.1088/1475-7516/2016/12/005},
  url = {https://dx.doi.org/10.1088/1475-7516/2016/12/005},
  year = {2016},
  volume = {2016},
  number = {12},
  pages = {005},
  author = {{Ming-Jian} Zhang and {Jun-Qing} Xia},
  title = {{Test of the cosmic evolution using Gaussian processes}},
  journal = {J. Cosmol. Astropart. Phys.}
}

@article{Wolf_2025,
   title={Assessing Cosmological Evidence for Nonminimal Coupling},
   volume={135},
   ISSN={1079-7114},
   url={http://dx.doi.org/10.1103/jysf-k72m},
   DOI={10.1103/jysf-k72m},
   number={8},
   journal={Physical Review Letters},
   publisher={American Physical Society (APS)},
   author={Wolf, William J. and García-García, Carlos and Anton, Theodore and Ferreira, Pedro G.},
   year={2025},
   month=Aug }

@article{Di_Valentino_2026,
   title={Tensions in Cosmology 2025},
   volume={10},
   ISSN={2397-3366},
   url={http://dx.doi.org/10.1038/s41550-026-02781-1},
   DOI={10.1038/s41550-026-02781-1},
   number={2},
   journal={Nature Astronomy},
   publisher={Springer Science and Business Media LLC},
   author={Di Valentino, Eleonora and Said, Jackson Levi and Saridakis, Emmanuel N.},
   year={2026},
   month=Feb, pages={180–182} }

@article{Kumar_Aluri_2023,
   title={Is the observable Universe consistent with the cosmological principle?},
   volume={40},
   ISSN={1361-6382},
   url={http://dx.doi.org/10.1088/1361-6382/acbefc},
   DOI={10.1088/1361-6382/acbefc},
   number={9},
   journal={Classical and Quantum Gravity},
   publisher={IOP Publishing},
   author={Kumar Aluri, Pavan and Cea, Paolo and Chingangbam, Pravabati and Chu, Ming-Chung and Clowes, Roger G and Hutsemékers, Damien and Kochappan, Joby P and Lopez, Alexia M and Liu, Lang and Martens, Niels C M and Martins, C J A P and Migkas, Konstantinos and Ó Colgáin, Eoin and Pranav, Pratyush and Shamir, Lior and Singal, Ashok K and Sheikh-Jabbari, M M and Wagner, Jenny and Wang, Shao-Jiang and Wiltshire, David L and Yeung, Shek and Yin, Lu and Zhao, Wen},
   year={2023},
   month=Apr, pages={094001} }

@article{BERNARDO2023101213,
title = {Dark energy by natural evolution: Constraining dark energy using Approximate Bayesian Computation},
journal = {Physics of the Dark Universe},
volume = {40},
pages = {101213},
year = {2023},
issn = {2212-6864},
doi = {https://doi.org/10.1016/j.dark.2023.101213},
url = {https://www.sciencedirect.com/science/article/pii/S221268642300047X},
author = {Reginald Christian Bernardo and Daniela Grandón and Jackson {Levi Said} and Víctor H. Cárdenas}
}

@article{tr6y-kpc6,
  title = {{DESI DR2 results. II. Measurements of baryon acoustic oscillations and cosmological constraints}},
  author = {Abdul Karim, M. and Aguilar, J. and Ahlen, S. and Alam and others},
  collaboration = {DESI Collaboration},
  journal = {Phys. Rev. D},
  volume = {112},
  issue = {8},
  pages = {083515},
  numpages = {40},
  year = {2025},
  month = {Oct},
  publisher = {American Physical Society},
  doi = {10.1103/tr6y-kpc6},
  url = {https://link.aps.org/doi/10.1103/tr6y-kpc6}
}

@article{Carroll_2004,
  title = {{Is cosmic speed-up due to new gravitational physics?}},
  author = {Carroll, Sean M. and Duvvuri, Vikram and Trodden, Mark and Turner, Michael S.},
  journal = {Phys. Rev. D},
  volume = {70},
  pages = {043528},
  year = {2004},
  doi = {10.1103/PhysRevD.70.043528},
  url = {https://link.aps.org/doi/10.1103/PhysRevD.70.043528}
}

@article{Naruko_2016,
   title={Gravitational scalar–tensor theory},
   volume={33},
   ISSN={1361-6382},
   url={http://dx.doi.org/10.1088/0264-9381/33/9/09LT01},
   DOI={10.1088/0264-9381/33/9/09lt01},
   number={9},
   journal={Classical and Quantum Gravity},
   publisher={IOP Publishing},
   author={Naruko, Atsushi and Yoshida, Daisuke and Mukohyama, Shinji},
   year={2016},
   month=Apr, pages={09LT01} }

@article{Riess_1998,
doi = {10.1086/300499},
url = {https://dx.doi.org/10.1086/300499},
year = {1998},
volume = {116},
pages = {1009},
author = {Adam G. Riess and Alexei V. Filippenko and Peter Challis and Alejandro Clocchiatti and others},
title = {{Observational Evidence from Supernovae for an Accelerating Universe and a Cosmological Constant}},
journal = {Astron. J.},
}

@article{PhysRevD.92.104042,
  title = {Modified teleparallel theories of gravity},
  author = {Bahamonde, Sebastian and B\"ohmer, Christian G. and Wright, Matthew},
  journal = {Phys. Rev. D},
  volume = {92},
  issue = {10},
  pages = {104042},
  numpages = {7},
  year = {2015},
  month = {Nov},
  publisher = {American Physical Society},
  doi = {10.1103/PhysRevD.92.104042},
  url = {https://link.aps.org/doi/10.1103/PhysRevD.92.104042}
}

@misc{Seikel2013,
  url = {https://arxiv.org/abs/1311.6678},
  year = {2013},
  author = {Marina Seikel and Chris Clarkson},
  title = {Optimising Gaussian processes for reconstructing dark energy dynamics from supernovae},
  archivePrefix = {arXiv},
  eprint = {1311.6678},
  primaryClass = {astro-ph.CO}
}

@article{Carroll_2001,
   title={The Cosmological Constant},
   volume={4},
   ISSN={1433-8351},
   url={http://dx.doi.org/10.12942/lrr-2001-1},
   DOI={10.12942/lrr-2001-1},
   number={1},
   journal={Living Reviews in Relativity},
   publisher={Springer Science and Business Media LLC},
   author={Carroll, Sean M.},
   year={2001},
   month=Feb }

@book{einstein1920relativity,
  author       = {Einstein, Albert},
  title        = {Relativity: The Special and General Theory},
  translator   = {Lawson, Robert W.},
  publisher    = {Methuen \& Co. Ltd.},
  year         = {1920},
  url          ={https://www.marxists.org/reference/archive/einstein/works/1910s/relative/relativity.pdf}
}

@book{lorentz2026principle,
  title={The Principle of Relativity},
  author={Lorentz, H. and Einstein, A. and Minkowski, H.},
  isbn={9781025922652},
  url={https://books.google.co.in/books?id=Kw_z0QEACAAJ},
  year={2026},
  publisher={Creative Media Partners, LLC}
}

@misc{efstathiou2024challengeslambdacdmcosmology,
      title={Challenges to the {$\Lambda$ CDM} Cosmology}, 
      author={George Efstathiou},
      year={2024},
      eprint={2406.12106},
      archivePrefix={arXiv},
      primaryClass={astro-ph.CO},
      url={https://arxiv.org/abs/2406.12106}, 
}

@InProceedings{10.1007/978-3-031-85527-6_13,
    author = "Perivolaropoulos, Leandros and Tsagas, Christos",
     title = "The Challenges for {$\Lambda $}{CDM} and the Physics Transition Approaches",
    booktitle = "A Multipolar Universe?",
    year = "2025",
    publisher = "Springer Nature Switzerland",
    pages = "181--251"
}

@ARTICLE{2017ARA&A..55..343B,
  author       = {{Bullock}, James S. and {Boylan-Kolchin}, Michael},
  title        = "{Small-Scale Challenges to the {$\Lambda$CDM} Paradigm}",
  journal      = {Annual Review of Astronomy and Astrophysics},
  year         = {2017},
  volume       = {55},
  number       = {1},
  pages        = {343--387},
  doi          = {10.1146/annurev-astro-091916-055313},
  archivePrefix = {arXiv},
  primaryClass = {astro-ph.CO},
  url          = {https://arxiv.org/abs/1707.04256}
}

@article{YADAV2026100453,
title = {Empirical validation: Investigating the {$\Lambda_s$CDM} model with new DESI BAO observations},
journal = {Journal of High Energy Astrophysics},
volume = {49},
pages = {100453},
year = {2026},
issn = {2214-4048},
doi = {https://doi.org/10.1016/j.jheap.2025.100453},
url = {https://www.sciencedirect.com/science/article/pii/S221440482500134X},
author = {Manish Yadav and Archana Dixit and Anirudh Pradhan and M.S. Barak}
}

@article{DiValentino_2021,
doi = {10.1088/1361-6382/ac086d},
url = {https://doi.org/10.1088/1361-6382/ac086d},
year = {2021},
month = {jul},
publisher = {IOP Publishing},
volume = {38},
number = {15},
pages = {153001},
author = {Di Valentino, Eleonora and Mena, Olga and Pan, Supriya and Visinelli, Luca and Yang, Weiqiang and Melchiorri, Alessandro and Mota, David F and Riess, Adam G and Silk, Joseph},
title = {In the realm of the Hubble tension—a review of solutions},
journal = {Classical and Quantum Gravity}
}

@article{Abdalla_2022,
   title={Cosmology intertwined: A review of the particle physics, astrophysics, and cosmology associated with the cosmological tensions and anomalies},
   volume={34},
   ISSN={2214-4048},
   url={http://dx.doi.org/10.1016/j.jheap.2022.04.002},
   DOI={10.1016/j.jheap.2022.04.002},
   journal={Journal of High Energy Astrophysics},
   publisher={Elsevier BV},
   author={Abdalla, Elcio and Abellán, Guillermo Franco and Aboubrahim, Amin and Agnello and others},
   year={2022},
   month={June}, pages={49–211} }

@misc{boiza2025addressingh0s8tensions,
      title={Addressing {$H_0$} and {$S_8$} tensions within {$f(Q)$} cosmology}, 
      author={Carlos G. Boiza and Maria Petronikolou and Mariam Bouhmadi-López and Emmanuel N. Saridakis},
      year={2025},
      eprint={2505.18264},
      archivePrefix={arXiv},
      primaryClass={astro-ph.CO},
      url={https://arxiv.org/abs/2505.18264}, 
}

@article{DESI:2025dr2,
  title = {{DESI DR2 results. II. Measurements of baryon acoustic oscillations and cosmological constraints}},
  author = {Abdul Karim, M. and Aguilar, J. and Ahlen, S. and Alam and others.},
  collaboration = {DESI Collaboration},
  journal = {Phys. Rev. D},
  volume = {112},
  issue = {8},
  pages = {083515},
  numpages = {40},
  year = {2025},
  month = {Oct},
  publisher = {American Physical Society},
  doi = {10.1103/tr6y-kpc6},
  url = {https://link.aps.org/doi/10.1103/tr6y-kpc6}
}

@article{DESI:2025lya,
  title = {{DESI DR2 results. I. Baryon acoustic oscillations from the Lyman alpha forest}},
  author = {Abdul Karim, M. and Aguilar, J. and Ahlen, S. and Allende Prieto and others},
  collaboration = {DESI Collaboration},
  journal = {Phys. Rev. D},
  volume = {112},
  issue = {8},
  pages = {083514},
  numpages = {28},
  year = {2025},
  month = {Oct},
  publisher = {American Physical Society},
  doi = {10.1103/2wwn-xjm5},
  url = {https://link.aps.org/doi/10.1103/2wwn-xjm5}
}

@article{Abdul_Karim_2025,
   title = "{DESI DR2 Results II: Measurements of Baryon Acoustic Oscillations and Cosmological Constraints}",
   volume = "{112}",
   ISSN = "{2470-0029}",
   url = "{http://dx.doi.org/10.1103/tr6y-kpc6}",
   DOI = "{10.1103/tr6y-kpc6}",
   number = "{8}",
   journal = "{Physical Review D}",
   publisher = "{American Physical Society (APS)}",
   author = "{Abdul Karim, M. and Aguilar, J. and Ahlen, S. and Alam and others}",
   year = "{2025}",
   month = Oct }

@book{Burnham2002,
  author    = {Burnham, Kenneth P. and Anderson, David R.},
  title     = {Model Selection and Multimodel Inference: A Practical Information-Theoretic Approach},
  edition   = {2nd},
  publisher = {Springer-Verlag},
  address   = {New York},
  year      = {2002},
  isbn      = {978-0-387-95364-9},
  doi       = {10.1007/b97636}
}

@article{Schwarz1978,
  author  = {Schwarz, Gideon},
  title   = {Estimating the Dimension of a Model},
  journal = {The Annals of Statistics},
  volume  = {6},
  number  = {2},
  pages   = {461--464},
  year    = {1978},
  doi     = {10.1214/aos/1176344136}
}

@article{Spiegelhalter2002,
  author  = {Spiegelhalter, David J. and Best, Nicola G. and Carlin, Bradley P. and van der Linde, Angelika},
  title   = {Bayesian Measures of Model Complexity and Fit},
  journal = {Journal of the Royal Statistical Society: Series B (Statistical Methodology)},
  volume  = {64},
  number  = {4},
  pages   = {583--639},
  year    = {2002},
  doi     = {10.1111/1467-9868.00353}
}

@article{KassRaftery1995,
  author  = {Kass, Robert E. and Raftery, Adrian E.},
  title   = {Bayes Factors},
  journal = {Journal of the American Statistical Association},
  volume  = {90},
  number  = {430},
  pages   = {773--795},
  year    = {1995},
  doi     = {10.1080/01621459.1995.10476572}
}

@article{Lewis2002,
  title = {{Cosmological parameters from CMB and other data: A Monte Carlo approach}},
  author = {Lewis, Antony and Bridle, Sarah},
  journal = {Phys. Rev. D},
  volume = {66},
  issue = {10},
  pages = {103511},
  numpages = {16},
  year = {2002},
  month = {Nov},
  publisher = {American Physical Society},
  doi = {10.1103/PhysRevD.66.103511},
  url = {https://link.aps.org/doi/10.1103/PhysRevD.66.103511}
}

@article{Handley2015,
    author = {Handley, W. J. and Hobson, M. P. and Lasenby, A. N.},
    title = {{polychord: next-generation nested sampling}},
    journal = {Monthly Notices of the Royal Astronomical Society},
    volume = {453},
    number = {4},
    pages = {4384-4398},
    year = {2015},
    month = {11},
    issn = {0035-8711},
    doi = {10.1093/mnras/stv1911},
    url = {https://doi.org/10.1093/mnras/stv1911},
}

@article{Lewis_2013,
   title = {Efficient sampling of fast and slow cosmological parameters},
   volume = {87},
   ISSN = {1550-2368},
   url = {http://dx.doi.org/10.1103/PhysRevD.87.103529},
   DOI = {10.1103/physrevd.87.103529},
   number = {10},
   journal = {Physical Review D},
   publisher = {American Physical Society (APS)},
   author = {Lewis, Antony},
   year = {2013},
   month = May }

@ARTICLE{Zhang2014,
       author = {{Zhang}, Cong and {Zhang}, Han and {Yuan}, Shuo and {Liu}, Siqi and {Zhang}, Tong-Jie and {Sun}, Yan-Chun},
        title = "{Four new observational H(z) data from luminous red galaxies in the Sloan Digital Sky Survey data release seven}",
      journal = {Research in Astronomy and Astrophysics},
         year = 2014,
        month = oct,
       volume = {14},
       number = {10},
          eid = {1221-1233},
        pages = {1221-1233},
          doi = {10.1088/1674-4527/14/10/002},
 primaryClass = {astro-ph.CO},
       adsurl = {https://ui.adsabs.harvard.edu/abs/2014RAA....14.1221Z}
}

@article{Simon_2005,
   title={Constraints on the redshift dependence of the dark energy potential},
   volume={71},
   ISSN={1550-2368},
   url={http://dx.doi.org/10.1103/PhysRevD.71.123001},
   DOI={10.1103/physrevd.71.123001},
   number={12},
   journal={Physical Review D},
   publisher={American Physical Society (APS)},
   author={Simon, Joan and Verde, Licia and Jimenez, Raul},
   year={2005},
   month= {June} }

@article{M_Moresco_2012,
   title={Improved constraints on the expansion rate of the Universe up to {$z \sim 1.1$} from the spectroscopic evolution of cosmic chronometers},
   volume={2012},
   ISSN={1475-7516},
   url={http://dx.doi.org/10.1088/1475-7516/2012/08/006},
   DOI={10.1088/1475-7516/2012/08/006},
   number={08},
   journal={Journal of Cosmology and Astroparticle Physics},
   publisher={IOP Publishing},
   author={M. Moresco and A. Cimatti and R. Jimenez and L. Pozzetti and G. Zamorani and M. Bolzonella and J. Dunlop and F. Lamareille and others},
   year={2012},
   month=Aug, 
   pages={006–006} }

@misc{neal2005takingbiggermetropolissteps,
      title={Taking Bigger Metropolis Steps by Dragging Fast Variables}, 
      author={Radford M. Neal},
      year={2005},
      eprint={math/0502099},
      archivePrefix={arXiv},
      primaryClass={math.ST},
      url={https://arxiv.org/abs/math/0502099}, 
}

@article{Handley_2015,
   title={polychord: next-generation nested sampling},
   volume={453},
   ISSN={1365-2966},
   url={http://dx.doi.org/10.1093/mnras/stv1911},
   DOI={10.1093/mnras/stv1911},
   number={4},
   journal={Monthly Notices of the Royal Astronomical Society},
   publisher={Oxford University Press (OUP)},
   author={Handley, W. J. and Hobson, M. P. and Lasenby, A. N.},
   year={2015},
   month={Sept}, pages={4385–4399} }

@misc{cartis2018improvingflexibilityrobustnessmodelbased,
      title={Improving the Flexibility and Robustness of Model-Based Derivative-Free Optimization Solvers}, 
      author={Coralia Cartis and Jan Fiala and Benjamin Marteau and Lindon Roberts},
      year={2018},
      eprint={1804.00154},
      archivePrefix={arXiv},
      primaryClass={math.OC},
      url={https://arxiv.org/abs/1804.00154}, 
}

@article{Cartis_2021,
   title={Escaping local minima with local derivative-free methods: a numerical investigation},
   volume={71},
   ISSN={1029-4945},
   url={http://dx.doi.org/10.1080/02331934.2021.1883015},
   DOI={10.1080/02331934.2021.1883015},
   number={8},
   journal={Optimization},
   publisher={Informa UK Limited},
   author={Cartis, Coralia and Roberts, Lindon and Sheridan-Methven, Oliver},
   year={2021},
   month=Feb, pages={2343–2373} }

@misc{lodha2025extendeddarkenergyanalysis,
      title = "{Extended Dark Energy analysis using DESI DR2 BAO measurements}", 
      author = "{K. Lodha and R. Calderon and W. L. Matthewson and A. Shafieloo and M. Ishak and J. Pan and C. Garcia-Quintero and D. Huterer and others}",
      year = "{2025}",
      eprint = "{2503.14743}",
      archivePrefix = "{arXiv}",
      primaryClass = "{astro-ph.CO}",
      url = "{https://arxiv.org/abs/2503.14743}" 
}

@article{Silva_2025,
   title= {Testing signatures of phantom crossing through full-shape galaxy clustering analysis},
   volume={2025},
   ISSN={1475-7516},
   url={http://dx.doi.org/10.1088/1475-7516/2025/11/078},
   DOI={10.1088/1475-7516/2025/11/078},
   number={11},
   journal={Journal of Cosmology and Astroparticle Physics},
   publisher={IOP Publishing},
   author={Silva, Emanuelly and Nunes, Rafael C.},
   year={2025},
   month=Nov, pages={078} }

@BOOK{Weinberg:2008,
       author = {{Weinberg}, Steven},
        title = "{Cosmology}",
        publisher = {Oxford University Press},
         year = 2008,
       adsurl = {https://ui.adsabs.harvard.edu/abs/2008cosm.book.....W}
}

@article{Weinberg:1988cp,
    author = "Weinberg, Steven",
    title = "{The Cosmological Constant Problem}",
    journal = "Rev. Mod. Phys.",
    volume = "61",
    pages = "1--23",
    year = "1989",
    doi = "10.1103/RevModPhys.61.1"
}

@article{Peebles_2003,
   title={The cosmological constant and dark energy},
   volume={75},
   ISSN={1539-0756},
   url={http://dx.doi.org/10.1103/RevModPhys.75.559},
   DOI={10.1103/revmodphys.75.559},
   number={2},
   journal={Reviews of Modern Physics},
   publisher={American Physical Society (APS)},
   author={Peebles, P. J. E. and Ratra, Bharat},
   year={2003},
   month=Apr, pages={559–606} }

@article{Perlmutter_1997,
doi = {10.1086/304265},
url = {https://dx.doi.org/10.1086/304265},
year = {1997},
volume = {483},
pages = {565},
author = {S. Perlmutter and S. Gabi and G. Goldhaber and A. Goobar and others},
title = {{Measurements of the Cosmological Parameters $\Omega$ and $\Lambda$ from the First Seven Supernovae at $z\geq 0.35$}},
journal = {Astrophys. J.}
}

@article{Perlmutter_1998,
doi = {10.1038/34124},
url = {https://doi.org/10.1038/34124},
year = {1998},
volume = {391},
pages = {51},
author = {S. Perlmutter and G. Aldering and M. Della Valle and S. Deustua and others},
title = {{Discovery of a supernova explosion at half the age of the Universe}},
journal = {nature}
}

@article{Perlmutter_1999,
doi = {10.1086/307221},
url = {https://dx.doi.org/10.1086/307221},
year = {1999},
volume = {517},
pages = {565},
author = {S. Perlmutter and G. Aldering and G. Goldhaber and R. A. Knop and others},
title = {{Measurements of $\Omega$ and $\Lambda$ from $42$ High-Redshift Supernovae}},
journal = {Astrophys. J.}
}

@article{Riess_2004,
doi = {10.1086/383612},
url = {https://dx.doi.org/10.1086/383612},
year = {2004},
volume = {607},
pages = {665},
author = {Adam G. Riess and Louis-Gregory Strolger and John Tonry and Stefano Casertano and others},
title = {{Type Ia Supernova Discoveries at $z > 1$ from the Hubble Space Telescope: Evidence for Past Deceleration and Constraints on Dark Energy Evolution}},
journal = {Astrophys. J.}
}

@article{Riess_2007,
doi = {10.1086/510378},
url = {https://dx.doi.org/10.1086/510378},
year = {2007},
volume = {659},
pages = {98},
author = {Adam G. Riess and Louis-Gregory Strolger and Stefano Casertano and Henry C. Ferguson and others},
title = {{New Hubble Space Telescope Discoveries of Type Ia Supernovae at $z\geq 1$: Narrowing Constraints on the Early Behavior of Dark Energy*}},
journal = {Astrophys. J.},
}

@article{Verde_2019,
   title={Tensions between the early and late Universe},
   volume={3},
   ISSN={2397-3366},
   url={http://dx.doi.org/10.1038/s41550-019-0902-0},
   DOI={10.1038/s41550-019-0902-0},
   number={10},
   journal={Nature Astronomy},
   publisher={Springer Science and Business Media LLC},
   author={Verde, Licia and Treu, Tommaso and Riess, Adam G.},
   year={2019},
   month={Sept}, pages={891–895} }

@article{Bennett_2003,
doi = {10.1086/377253},
url = {https://dx.doi.org/10.1086/377253},
year = {2003},
volume = {148},
pages = {1},
author = {C. L. Bennett and M. Halpern and G. Hinshaw and N. Jarosik and others},
title = {{First-Year Wilkinson Microwave Anisotropy Probe
(WMAP)* Observations: Preliminary Maps and Basic Results}},
journal = {Astrophys. J. Supp. Ser.}
}

@article{Spergel_2003,
doi = {10.1086/377226},
url = {https://dx.doi.org/10.1086/377226},
year = {2003},
volume = {148},
pages = {175},
author = {D. N. Spergel and L. Verde and H. V. Peiris and E. Komatsu and others},
title = {{First-Year Wilkinson Microwave Anisotropy Probe
(WMAP)* Observations: Determination of Cosmological Parameters}},
journal = {Astrophys. J. Supp. Ser.}
}

@article{Clifton_2012,
   title={Modified gravity and cosmology},
   volume={513},
   ISSN={0370-1573},
   url={http://dx.doi.org/10.1016/j.physrep.2012.01.001},
   DOI={10.1016/j.physrep.2012.01.001},
   number={1-3},
   journal={Physics Reports},
   publisher={Elsevier BV},
   author={Clifton, Timothy and Ferreira, Pedro G. and Padilla, Antonio and Skordis, Constantinos},
   year={2012},
   month=Mar, pages={1–189} }

@article{Nojiri_2017,
   title={Modified gravity theories on a nutshell: Inflation, bounce and late-time evolution},
   volume={692},
   ISSN={0370-1573},
   url={http://dx.doi.org/10.1016/j.physrep.2017.06.001},
   DOI={10.1016/j.physrep.2017.06.001},
   journal={Physics Reports},
   publisher={Elsevier BV},
   author={Nojiri, S. and Odintsov, S.D. and Oikonomou, V.K.},
   year={2017},
   month={June}, pages={1–104} }

@article{Koyama_2016,
   title={Cosmological tests of modified gravity},
   volume={79},
   ISSN={1361-6633},
   url={http://dx.doi.org/10.1088/0034-4885/79/4/046902},
   DOI={10.1088/0034-4885/79/4/046902},
   number={4},
   journal={Reports on Progress in Physics},
   publisher={IOP Publishing},
   author={Koyama, Kazuya},
   year={2016},
   month={Mar}, pages={046902} }

@article{De_Felice_2010,
   title={${f(R)}$ Theories},
   volume={13},
   ISSN={1433-8351},
   url={http://dx.doi.org/10.12942/lrr-2010-3},
   DOI={10.12942/lrr-2010-3},
   number={1},
   journal={Living Reviews in Relativity},
   publisher={Springer Science and Business Media LLC},
   author={De Felice, Antonio and Tsujikawa, Shinji},
   year={2010},
   month={June} }

@article{Sotiriou_2010,
  author    = {Sotiriou, Thomas P. and Faraoni, Valerio},
  title     = {${f(R)}$ theories of gravity},
  journal   = {Reviews of Modern Physics},
  volume    = {82},
  number    = {1},
  pages     = {451--497},
  year      = {2010},
  month     = mar,
  publisher = {American Physical Society},
  doi       = {10.1103/RevModPhys.82.451},
  url       = {https://doi.org/10.1103/RevModPhys.82.451}
}

@book{faraoni2004cosmology,
  title={Cosmology in Scalar-Tensor Gravity},
  author={Faraoni, Valerio},
  year={2004},
  publisher={Kluwer Academic Publishers},
  address={Dordrecht},
  series={Fundamental Theories of Physics},
  volume={139},
  doi={10.1007/978-1-4020-1989-0}
}

@article{Cai:2015emx,
    author = "Cai, Yi-Fu and Capozziello, Salvatore and De Laurentis, Mariafelicia and Saridakis, Emmanuel N. Saridakis",
    title = "{Teleparallel gravity: from theory to cosmology}",
    archivePrefix = "arXiv",
    primaryClass = "gr-qc",
    doi = "10.1088/0034-4885/79/10/106901",
    journal = "Rept. Prog. Phys.",
    volume = "79",
    number = "10",
    pages = "106901",
    year = "2016"
}

@article{Harko_2012,
   title={Metric-Palatini gravity unifying local constraints and late-time cosmic acceleration},
   volume={85},
   ISSN={1550-2368},
   url={http://dx.doi.org/10.1103/PhysRevD.85.084016},
   DOI={10.1103/physrevd.85.084016},
   number={8},
   journal={Physical Review D},
   publisher={American Physical Society (APS)},
   author={Harko, Tiberiu and Koivisto, Tomi S. and Lobo, Francisco S. N. and Olmo, Gonzalo J.},
   year={2012},
   month=Apr }

@article{Harko_2014,
   title={Thermodynamic interpretation of the generalized gravity models with geometry-matter coupling},
   volume={90},
   ISSN={1550-2368},
   url={http://dx.doi.org/10.1103/PhysRevD.90.044067},
   DOI={10.1103/physrevd.90.044067},
   number={4},
   journal={Physical Review D},
   publisher={American Physical Society (APS)},
   author={Harko, Tiberiu},
   year={2014},
   month=Aug }

@article{Wang_2012,
  author    = {Wang, Jun and Liao, Kai},
  title     = {Energy conditions in ${f(R,L_m)}$ gravity},
  journal   = {Classical and Quantum Gravity},
  volume    = {29},
  number    = {21},
  pages     = {215016},
  year      = {2012},
  month     = oct,
  publisher = {IOP Publishing},
  doi       = {10.1088/0264-9381/29/21/215016},
  url       = {https://doi.org/10.1088/0264-9381/29/21/215016}
}

@article{Avelino:2018rsb,
    author = "Avelino, P. P. and Azevedo, R. P. L.",
    title = "{Perfect fluid Lagrangian and its cosmological implications in theories of gravity with nonminimally coupled matter fields}",
    archivePrefix = "arXiv",
    primaryClass = "gr-qc",
    doi = "10.1103/PhysRevD.97.064018",
    journal = "Phys. Rev. D",
    volume = "97",
    number = "6",
    pages = "064018",
    year = "2018"
}

@misc{gonçalves2024causalityviolationfrlmtgravity,
      title={On causality and its violation in {$f(R,Lm,T)$} gravity}, 
      author={J. S. Gonçalves and A. F. Santos},
      year={2024},
      eprint={2411.11588},
      archivePrefix={arXiv},
      primaryClass={gr-qc},
      url={https://arxiv.org/abs/2411.11588}, 
}

@article{DEVI2024101640,
  doi = {https://doi.org/10.1016/j.dark.2024.101640},
  url = {https://www.sciencedirect.com/science/article/pii/S221268642400222X},
  year = {2024},
  volume = {46},
  pages = {101640},
  author = {Y. Kalpana Devi and S.A. Narawade and B. Mishra},
  title = {Constraining parameters for the accelerating universe in {$f(R,Lm)$} gravity},
  journal = {Physics of the Dark Universe}
}

@article{DEVI2026170597,
  title = {Accelerating cosmological model with scalar field in {$f(R,Lm)$} gravity},
  journal = {Annals of Physics},
  volume = {493},
  pages = {170597},
  year = {2026},
  issn = {0003-4916},
  doi = {https://doi.org/10.1016/j.aop.2026.170597},
  url = {https://www.sciencedirect.com/science/article/pii/S0003491626002563},
  author = {Y. Kalpana Devi and Rahul Bhagat and B. Mishra}
}

@article{Kass01061995,
author = {Robert E. Kass and Adrian E. Raftery},
title = {Bayes Factors},
journal = {Journal of the American Statistical Association},
volume = {90},
number = {430},
pages = {773--795},
year = {1995},
publisher = {Taylor \& Francis},
doi = {10.1080/01621459.1995.10476572},
URL = {https://doi.org/10.1080/01621459.1995.10476572},
eprint ={https://doi.org/10.1080/01621459.1995.10476572}}

@book{Jeffreys:1939xee,
    author = "{Jeffreys, Harold}",
    title = "{The Theory of Probability}",
    publisher = {Oxford University Press},
    isbn = "978-0-19-850368-2, 978-0-19-853193-7",
    series = "Oxford Classic Texts in the Physical Sciences",
    year = "1939"
}

@article{Liddle_2007,
   title={Information criteria for astrophysical model selection},
   volume={377},
   ISSN={1745-3925},
   url={http://dx.doi.org/10.1111/j.1745-3933.2007.00306.x},
   DOI={10.1111/j.1745-3933.2007.00306.x},
   number={1},
   journal={Monthly Notices of the Royal Astronomical Society: Letters},
   publisher={Oxford University Press (OUP)},
   author={Liddle, Andrew R.},
   year={2007},
   month=May, pages={L74–L78} }

@article{Riess_2022,
  doi = {10.3847/2041-8213/ac5c5b},
  url = {https://doi.org/10.3847/2041-8213/ac5c5b},
  year = {2022},
  month = {jul},
  publisher = {The American Astronomical Society},
  volume = {934},
  number = {1},
  pages = {L7},
  author = {Riess, Adam G. and Yuan, Wenlong and Macri, Lucas M. and Scolnic, Dan and Brout, Dillon and Casertano, Stefano and Jones, David O. and Murakami, Yukei and Anand, Gagandeep S. and Breuval, Louise and Brink, Thomas G. and Filippenko, Alexei V. and Hoffmann, Samantha and Jha, Saurabh W. and D'arcy Kenworthy, W. and Mackenty, John and Stahl, Benjamin E. and Zheng, WeiKang},
  title = {A Comprehensive Measurement of the Local Value of the Hubble Constant with 1 km s$^{-1}$ Mpc$^{-1}$ Uncertainty from the Hubble Space Telescope and the SH0ES Team},
  journal = {The Astrophysical Journal Letters}
}

\end{document}